%% file: main.tex
\documentclass[letterpaper,twocolumn,10pt]{article}
\usepackage{usenix2020_SOUPS}

\usepackage{graphicx}

\usepackage{tikz}
\usetikzlibrary{shapes.geometric, arrows, positioning}
\usepackage{amsfonts, amsmath,amssymb}
\usetikzlibrary{arrows}
\usepackage{float}

\usepackage{tikz}
\usepackage{xspace}
\usepackage[utf8]{inputenc}
\usetikzlibrary{matrix}
\usepackage{amsmath}
\usepackage{cite}
\usepackage{tabularx}
 \usepackage{multirow}
 \usepackage{makecell}
\usepackage{booktabs}
\usepackage{graphicx,wrapfig,lipsum}
\usepackage{url}

\usepackage{caption}
 \usepackage{subcaption}

\newcommand{\new}[1]{\textcolor{black}{#1}}

\newcommand{\x}{\mathbf{x}}
\newcommand{\ks}{\mathbb{X}}
\newcommand{\ds}{\mathbf{X}} %train-set
\newcommand{\sem}[1]{\textbf{\{#1\}}}
\newcommand{\ppp}[1]{\scriptsize{#1}}
\newcommand{\rc}{\bullet} %removed character

\newcommand{\T}[1]{\textit{\lq\text{#1}\rq}}
\newcommand{\TT}[1]{``\textit{#1}"}
\newcommand{\ie}{{i.e.,~}}
\newcommand{\sota}{{state-of-the-art }}
\newcommand{\eg}{{e.g.,~}}

\newcommand{\etal}{{et~al.}} 

\newcommand{\refeq}[1]{Eq.~\ref{#1}}
\newcommand{\reffig}[1]{Figure~\ref{#1}}

\newcommand\myeq{\mkern1.5mu{=}\mkern1.5mu}
\newcommand{\rcc}{$\bullet$} %removed character

\newcommand{\BC}{X_{\text{BC}}}

\DeclareMathOperator*{\argmax}{arg\,max}
\DeclareMathOperator*{\argmin}{arg\,min}

\newcommand{\cd}[4]{\texttt{cov1d}[#1, #2, \textit{#3}, \textit{#4}] }
\newcommand{\rb}[2]{\texttt{ResblockBneck1D}[$\text{fn}\myeq#1$, $\text{ks}\myeq#2$] }

\begin{document}

%don't want date printed
\date{}

\title{\Large \bf Interpretable Probabilistic Password Strength Meters via Deep Learning}

% if you leave this blank it will default to a possibly ugly attempt 
% to make the contents of the \author command below into a string
\def\plainauthor{ Dario Pasquini et al.}

\author{
	{\rm Dario Pasquini}\\
	Stevens Institute of Technology, USA\\
	Sapienza University of Rome, Italy\\
	Institute of Applied Computing, CNR, Italy
	\and
	{\rm Giuseppe Ateniese}\\
	Stevens Institute of Technology, USA
	\and
	{\rm Massimo Bernaschi}\\
	Institute of Applied Computing, CNR, Italy
	% copy the following lines to add more authors
	% \and
	% {\rm Name}\\
	%Name Institution
} % end author
\maketitle
\footnotetext[1]{An abridged version of this paper appears in the proceedings of the 25th European Symposium on Research in Computer Security (ESORICS) 2020.}%
%
%-------------------------------------------------------------------------------
\begin{abstract}
	Probabilistic password strength meters have been proved to be the most accurate tools to measure password strength. Unfortunately, by construction, they are limited to solely produce an opaque security estimation that fails to fully support the user during the password composition. In the present work, we move the first steps towards cracking the intelligibility barrier of this compelling class of meters. We show that probabilistic password meters inherently own the capability to describe the latent relation between password strength and password structure. In our approach, the security contribution of each character composing a password is disentangled and used to provide explicit fine-grained feedback for the user.  
Furthermore, unlike existing heuristic constructions, our method is free from any human bias, and, more importantly, its feedback has a probabilistic interpretation.\\
In our contribution: (1) we formulate interpretable probabilistic password strength meters; (2) we describe how they can be implemented via an efficient and lightweight deep learning framework suitable for client-side operability.

\end{abstract}
\begin{figure}[t]
	\centering
	\begin{tikzpicture}
	\matrix[matrix of nodes, row sep=-.02\textwidth]{
		\includegraphics[width=.29\textwidth]{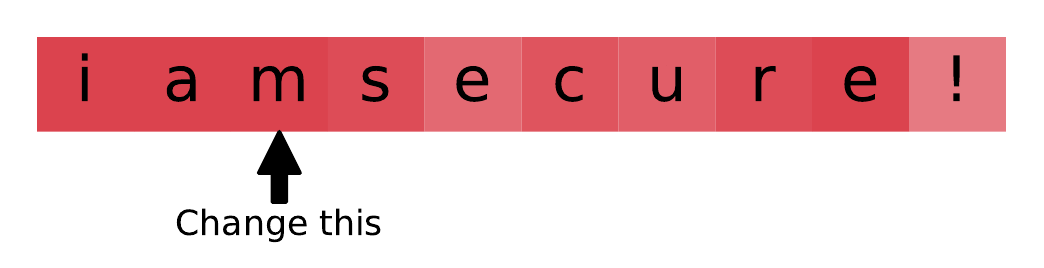} \\
		\includegraphics[width=.29\textwidth]{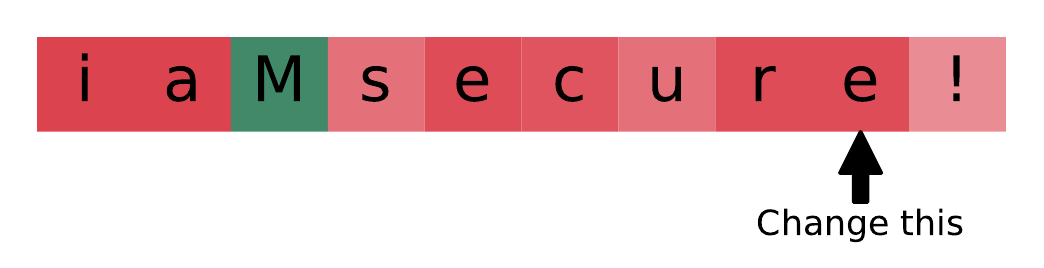} \\
		\includegraphics[width=.29\textwidth]{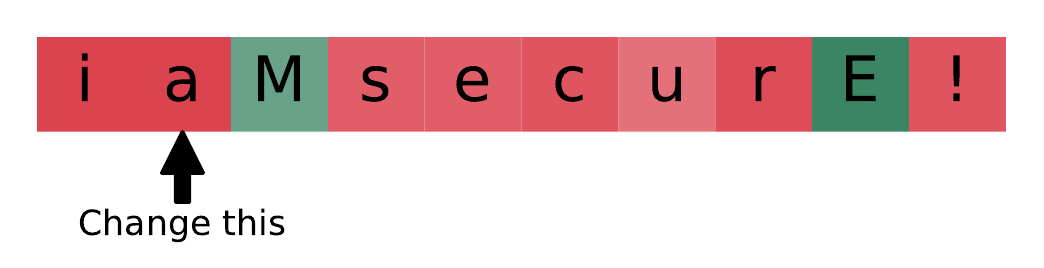} \\
		\includegraphics[width=.29\textwidth]{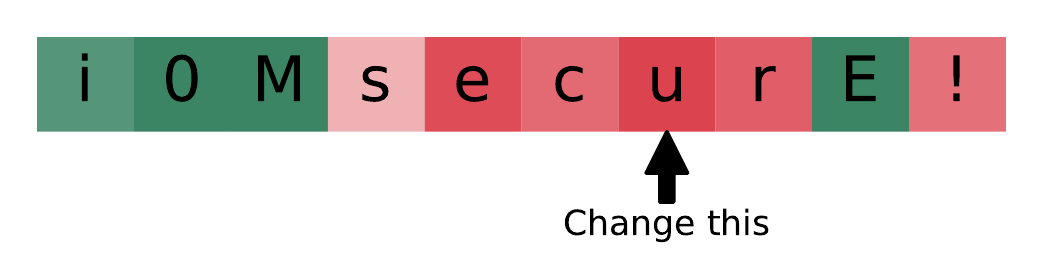} \\
		\includegraphics[width=.29\textwidth]{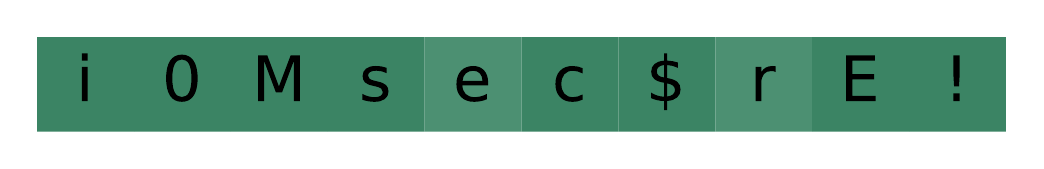} \\
	};
	\end{tikzpicture}
	\caption{Example of the character-level feedback mechanism and password composition process induced by our meter. In the figure, \TT{iamsecure!} is the password initially chosen by the user. Colors indicate the estimated character security: red~(insecure)~$\rightarrow$ green (secure).}
	\label{fig:per_ex}
\end{figure}
\section{Introduction}
%\todo{\lipsum[3-9]}\\
\input{intro}
%\new{\textbf{Organization:} Section~\ref{sec:intro} gives an overview of the fundamental concepts related to our contribution along with a brief survey of previous work. In Section~\ref{sec:meter}, ...	 defining the theoretical foundations behind interpretable PPSMs. Next, in Section~\ref{section:dp_impl}, we describe how such meter can be constructed relying on deep learning techniques. In Section~\ref{section:results}, we empirically validate the proposed probability estimation process as well as its deep learning implementation. Section~\ref{sec:conclusion} concludes the paper, although supplementary information are provided in the Appendices.}
%
%
\section{Background and preliminaries}
\label{sec:intro}
In this section, we offer an overview of the fundamental concepts that are important to understand our contribution. Section~\ref{ppsm_intro} covers Probabilistic Password Strength Meters. 
%Section~\ref{sec:nn} glances neural networks and related topics. 
Next, in Section~\ref{sec:spm}, we cover structured probabilistic models that will be fundamental in the interpretation of our approach. Finally, Section~\ref{sec:related} briefly discusses relevant previous works within the PSMs context.
\subsection{Probabilistic Password Strength Meters (PPSMs) }
\label{ppsm_intro}
Probabilistic password strength meters are PSMs that base their strength measure on an explicit estimate of password probability. In the process, they resort to probabilistic models to approximate the probability distribution behind a set of known passwords, typically, instances of a password leak. Having an approximation of the mass function, strength estimation is then derived by leveraging adversarial reasoning. Here, password robustness is estimated in consideration of an attacker who knows the underlying password distribution, and that aims at minimizing the guess entropy \cite{gue_and_en} of her/his guessing attack. To that purpose, the attacker performs an optimal guessing attack, where guesses are issued in decreasing probability order (\ie high-probability passwords first).
More formally, given a probability mass function $P(\x)$ defined on the key-space $\ks$, the attacker creates an ordering $\ks_{P(\x)}$ of $\ks$ such that:
\begin{equation}
\small
\ks_{P(\x)}=[x^0, x^1,\dots, x^n] \quad \text{where} \quad \forall_{i \in [0,n]}: P(x^i) \geq P(x^{i+1}) \quad .
\end{equation}
During the attack, the adversary produces guesses by traversing the list $\ks_{P(\x)}$. Under this adversarial model, passwords with high probability are considered weak, as they will be quickly guessed. Low-probability passwords, instead, are assessed as secure, as they will be matched by the attacker only after a considerable, possibly not feasible, number of guesses.
\vspace{-5pt}
\subsection{Structured Probabilistic Models}
\label{sec:spm}
Generally, the probabilistic models used by PPSMs are \textbf{probabilistic structured models} (even known as graphical models). These describe password distributions by leveraging a \textbf{graph notation} to illustrate the dependency properties among a set of random variables. Here, a random variable $\x_i$ is depicted as a vertex, and an edge between $\x_i$ and $\x_j$ exists whether $\x_i$ and $\x_j$ are statistically dependent. Structured probabilistic models are classified according to the orientation of edges. A direct acyclic graph (DAG) defines a \textbf{directed graphical model} (or Bayesian Network). In this formalism, an edge asserts a cause-effect relationship between two variables; that is, the state assumed from the variable $x_i$ is intended as a direct consequence of those assumed by its parents $par(\x_i)$ in the graph. Under such a description, a topological ordering among all the random variables can be asserted and used to factorize the joint probability distribution of the random variables effectively. On the other hand, an undirected graph defines an \textbf{undirected graphical model}, also known as Markov Random Field (MRF). In this description, the topological order among edges is relaxed, and connected variables influence each other symmetrically. However, this comes at the cost of giving up to any simple form of factorization of the joint distribution.

\subsection{Related Works}
\label{sec:related}
	\input{related}
\section{Meter foundations}
\label{sec:meter}
	In this section, we introduce the theoretical intuition behind the proposed meter. % as well as of the character-level feedback mechanism deriving from it.
	First, in Section~\ref{section:feedback_def}, we introduce and motivate the probabilistic character-level feedback mechanism. Later, in Section~\ref{section:mrf_def}, we describe how that mechanism can be obtained using undirected probabilistic models.
\begin{figure*}[t!]
	\centering
	\resizebox{1\textwidth}{!}{%
		\begin{tabular}{cccc}
			\begin{subfigure}{0.15\textwidth}\centering\includegraphics[scale=0.33]{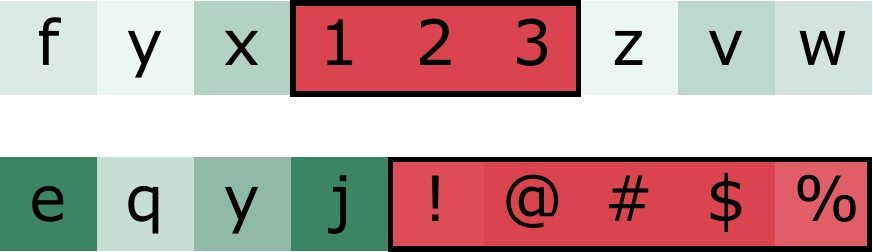}\caption{\tiny{Common tokens.}}\label{fig:taba}\end{subfigure}&
			\begin{subfigure}{0.15\textwidth}\centering\includegraphics[scale=0.33]{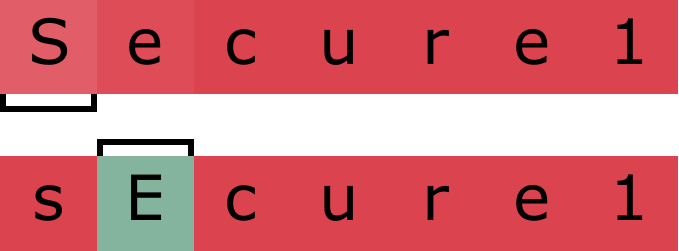}\caption{\tiny{Capitalize first/inner.}}\label{fig:tabb}\end{subfigure}&
			\begin{subfigure}{0.15\textwidth}\centering\includegraphics[scale=0.33]{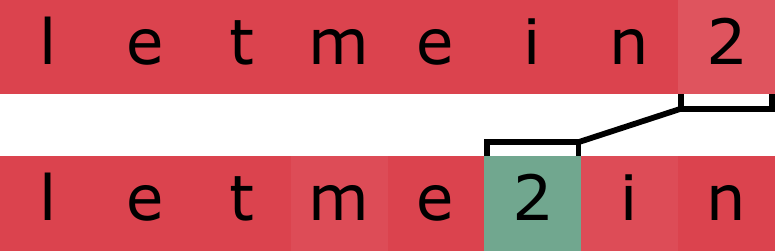}\caption{\tiny{Numeric last/inner}.}\label{fig:tabc}\end{subfigure}&
			\begin{subfigure}{0.15\textwidth}\centering\includegraphics[scale=0.33]{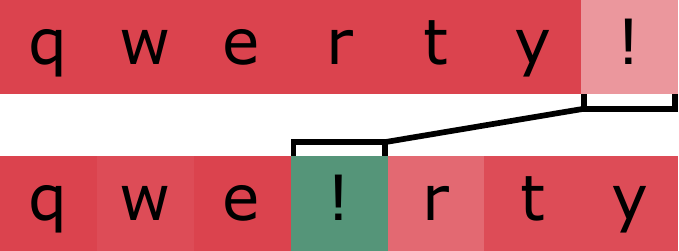}\caption{\tiny{Special last/inner.}}\label{fig:tabd}\end{subfigure}\\
		\end{tabular}
	}
	\caption{In Panel (a), the model automatically highlights the presence of weak substrings by assigning high probabilities to the characters composing them. Panels (b), (c), and (d) are examples of self-learned weak/strong password composition patterns. In panel (b), the model assigns a high probability to the capitalization of the first letter (a common practice), whereas it assigns low probability when the capitalization is performed on inner characters. Panel (c) and (d) report similar results for numeric and special characters.}
	\label{fig:self_feed}
\end{figure*}
\vspace{-5pt}
\subsection{Character-level strength estimation via probabilistic reasoning}
\label{section:feedback_def}
As introduced in Section \ref{ppsm_intro}, PPSMs employ probabilistic models to approximate the probability mass function of an observed password distribution, say $P(\x)$. Estimating $P(\x)$, however, could be particularly challenging, and suitable estimation techniques must be adopted to make the process feasible. In this direction, a general solution is to factorize the domain of the mass function (\ie the key-space); that is, passwords are modeled as a concatenation of smaller factors, typically, decomposed at the character level. Afterward, password distribution is estimated by modeling stochastic interactions among these simpler components.
More formally, every password is assumed as a realization $x=[x_1, \dots, x_{\ell}]$ of a random vector of the kind $\x=[\x_1, \dots, \x_{\ell}]$, where each disjoint random variable $\x_i$ represents the character at position $i$ in the string. Then, $P(\x)$ is described through \textbf{structured probabilistic models} that formalize the relations among those random variables, eventually defining a joint probability distribution. In the process, every random variable is associated with a \textbf{local conditional probability distribution} (here, referred to as $Q$) that describes the stochastic behavior of $\x_i$ in consideration of the conditional independence properties asserted from the underlying structured model \ie $Q(\x_i) \myeq P(\x_i \mid par(\x_i))$. Eventually, the joint measurement of probability is derived from the aggregation of the marginalized local conditional probability distributions, typically under the form $P(\x) \myeq \prod_{i=1}^{\ell} Q(\x_i \myeq x_i)$.\\ %For instance, the Markov model approach \cite{mm_first, MM} factorizes the joint probability as $P(\x) \myeq \prod_{i=1}^{\ell} P(\x_i | \x_{i-n},\dots, \x_{i-1})$, where $n$ is the order of the believed Markov property.\\
%\vspace{-5pt}

	As introduced in Section \ref{ppsm_intro}, the joint probability can be employed as a good representative for password strength. However, such a global assessment unavoidably hides much fine-grained information that can be extremely valuable to a password meter. In particular, the joint probability offers us an atomic interpretation of the password strength, but it fails at disentangling the relation between password strength and password structure. That is, it does not clarify which factors of an evaluated password are making that password insecure. However, as widely demonstrated by non-probabilistic approaches \cite{zxcvbn, FLA2, telepathwords}, users benefit from the awareness of which part of the chosen password is easily predictable and which is not. In this direction, we argue that the \textbf{local conditional probabilities} that naturally appear in the estimation of the joint one, if correctly shaped, can offer detailed insights into the strength or the weakness of each factor of a password.
	%These can act as the foundations to establish an interpretable probabilistic password strength meter capable of guiding users to the composition of secure as usable passwords.
\textbf{Such character-level probability assignments are an explicit interpretation of the relation between the structure of a password and its security.}
%These values can actively and fundamentally describe the strength or the weakness of each element composing the password}.
The main intuition here is that: high values of $Q(x_i)$ tell us that $x_i$ (\ie the character at position $i$ in the string) has a high impact on increasing the password probability and must be changed to make the password stronger. Instead, characters with low conditional probability are pushing the password to have low probability and must be maintained unchanged.
Figure~\ref{fig:self_feed} reports some visual representations of such probabilistic reasoning. Each segment's background color renders the value of the local conditional probability of the character. Red describes high probability values, whereas green describes low probability assignments.
	%This fine-grained estimation can be used as a sound guide for the user at composition time, taking the form of active feedback. 
	Such a mechanism can naturally discover weak passwords components and explicitly guide the user to explore alternatives.
	For instance, local conditional probabilities can spot the presence of predictable tokens in the password without the explicit use of dictionaries (\reffig{fig:taba}). These measurements are able to automatically describe common password patterns like those manually modeled from other approaches \cite{FLA2}, see Figures \ref{fig:tabb}, \ref{fig:tabc} and \ref{fig:tabd}. More importantly, they can potentially describe latent composition patterns that have never been observed and modeled by human beings. In doing this, neither supervision nor human-reasoning is required.
	\par
	
	Unfortunately, existing PPSMs, by construction, leverage arbitrary designed structured probabilistic models that make inefficient to produce the required estimates. Hereafter, we show that reshaping the mass estimation process will allow us to implement the feedback mechanism described above. To that purpose, we have to build a new probabilistic estimation framework and simulate a complete, undirected models.
\vspace{-5pt}
\subsection{An undirected description of password distribution} 
\label{section:mrf_def}
To simplify our method's understanding, we start with a description of the probabilistic reasoning of previous approaches. Then, we fully motivate our solution by comparison with them. In particular, we chose the \sota neural approach proposed in \cite{FLA} (henceforth, referred to as FLA) as a representative instance, since it is the least biased as well as the most accurate among the existing PPSMs.\\
FLA uses a recurrent neural network (RNN) to estimate password mass function at the character level. That model is \textbf{autoregressive} and assumes a stochastic process represented by a Bayesian network like the one depicted in \reffig{fig:gma}.
\input{gm}
\new{
%As previously anticipated, such a density estimation process bears the burden of bold assumptions on its formalization. 
 The description derived from a Bayesian Network implies a topological order among password characters. Here, characters influence their  local conditional  probabilities only asymmetrically; that is, the probability of $\x_{i+1}$ is conditioned by $\x_i$ but not \textit{vice versa}. In practice, for the  local conditional probabilities, this implies that the observation of the value assumed from $\x_{i+1}$ does not affect our belief in the value expected from $\x_i$, yet the opposite does. }
%However, this assumption is unrealistic and the underlying stochastic process may underestimate the single local conditional probabilities and the joint one.
The autoregressive nature of this model eventually simplifies the joint probability estimation; the local conditional probability of each character can be easily computed as $Q(\x_i)~\myeq~P(\x_i \mid \x_1, \dots, \x_{i-1})$, where $Q(\x_i)$ explicates that the $i$'th character solely depends on the characters that precede it in the string. Just as easily, the joint probability factorizes in:\[P(\x)~\myeq~\prod_{i\myeq1}^{\ell} Q(\x_i)~\myeq~P(\x_1) \prod_{i\myeq2}^\ell P(\x_i~\mid~\x_1 ,\dots \x_{i-1})\] by~chain~rule.\\
 
Unfortunately, although this approach does simplify the estimation process, the conditional probability $Q(x_i)$, \textit{per se}, does not provide a direct and coherent estimation of the security contribution of the single character $x_i$ in the password. This is particularly true for characters in the first positions of the string, even more so for the first character $x_1$, which is assumed to be independent of any other symbol in the password; its probability is the same for any possible configuration of the remaining random variables $[x_2,\dots,x_\ell]$. Nevertheless, in the context of a \textbf{sound} character-level feedback mechanism, the symbol $x_i$ must be defined as ``weak'' or ``strong'' according to the complete context defined by the entire string. For instance, given two passwords $y\myeq\TT{aaaaaaa}$ and $z\myeq\TT{a\scriptsize{\#\#\#\#\#\#}}$, the probability $Q(\x_1\myeq\T{a})$ should be different if measured on $y$ or $z$. More precisely, we expect $Q(\x_1\myeq\T{a}|y)$ to be much higher than $Q(\x_1\myeq\T{a}|z)$, as observing $y_{2,7}\myeq\TT{aaaaaa}$ drastically changes our expectations about the possible values assumed from the first character in the string. On the other hand, observing $z_{2,7} \myeq\TT{\scriptsize{\#\#\#\#\#\#}}$ tells us little about the event $\x_1\myeq\T{a}$. Yet, this interaction cannot be described through the Bayesian network reported in \reffig{fig:gma}, where $Q(\x_1\myeq\T{a}|y)$ eventually results equal to $Q(\x_1\myeq\T{a}|z)$. The same reasoning applies to trickier cases, as for the password $x\myeq\TT{(password)}$. Here, arguably, the security contribution of the first character $\T{(}$ strongly depends from the presence or absence of the last character\footnote{Even if not so common, strings enclosed among brackets or other special characters often appear in password leaks.}, \ie $x_7\myeq\T{)}$. The symbol $x_1\myeq\T{(}$, indeed, can be either a good choice (as it introduces entropy in the password) or a poor one (as it implies a predictable template in the password), but this solely depends on the value assumed from another character in the string (the last one in this example). 
%We argue that a good meter should be able to model similar templates and encourage the user to break them at composition time (\eg by advising the user to remove the first or the last bracket). Yet, again, such relation is \textit{a priori} excluded from existing structured models.\\
It should be apparent that the autoregressive nature of the model prevents the resulting local conditional probabilities from being sound descriptors of the real character probability as well as of their security contribution. Consequently, such measures cannot be used to build the feedback mechanism suggested in Section \ref{section:feedback_def}. The same conclusion applies to other classes of PPSMs \cite{MM, PCFG} which add even more structural biases on top of those illustrated by the model in \reffig{fig:gma}.\\

\iffalse
Under a broader view, directed models are intended to be used in contexts where the causality relationships among random variables can be fully understood. Unfortunately, even if passwords are physically written character after character by the users, it is impossible to assert neither independence nor cause-effect relations among the symbols that compose them. \textbf{Differently from plain dictionary words, passwords are built on top of much more complex structures and articulated interactions among characters that cannot be adequately described without relaxing many of the assumptions leveraged by existing PPSMs.} In the act of relaxing such assumptions, 
\fi
Differently, we base our estimation on an \textbf{undirected and complete\footnote{``Complete" in the graph-theory sense.} graphical model}, as this represents a  handier description of the password generative distribution.
% That is, neither independence nor causality among random variables are \textit{a priori} implied.
\reffig{fig:gmb} depicts the respective Markov Random Field (MRF) for passwords of length four. According to that description, the local conditional probability of the character $x_i$ directly depends on any other character in the string, \ie the full context. In other words, we model each variable $\x_i$ as a stochastic function of all the others. This intuition is better captured from the evaluation of local conditional probability (\refeq{eq:q}).
\begin{equation}
\label{eq:q}
Q(\x_i) =
\begin{cases}
P(\x_i \mid \x_{i+1}, \dots \x_{\ell}) & i=1\\
P(\x_i \mid \x_1,\dots \x_{i-1}) & i=\ell\\
P(\x_i \mid \x_1,\dots,\x_{i-1},\x_{i+1}, \dots \x_{\ell}) & 1<i<\ell \quad .
\end{cases}
\end{equation}
%
%The measurement $Q$ asserts that the probability of a character is potentially influenced by the configuration of all the other nodes in the graph. Nevertheless, if needed, independence or causality relations can be autonomously ascribed as context-specific independence\footnote{ independence that is verified only in case of specific configurations of the random variables.} from the probabilistic model itself. That is, the model is free to capture independence from the observed data and assert them as true at inference time.
Henceforth, we use the notation $Q(\x_i)$ to refer to the local conditional distribution of the $i$'th character within the password $x$. When $x$ is not clear from the context, we write $Q(\x_i\mid x)$ to make it explicit. The notation $Q(\x_i\myeq s)$ or $Q(s)$, instead, refers to the marginalization of the distribution according to the symbol~$s$.
\begin{figure}[t!]
		\resizebox{0.9\columnwidth}{!}{%
		\begin{tabular}{cc}
			\begin{subfigure}{0.4\columnwidth}\includegraphics[scale=0.4]{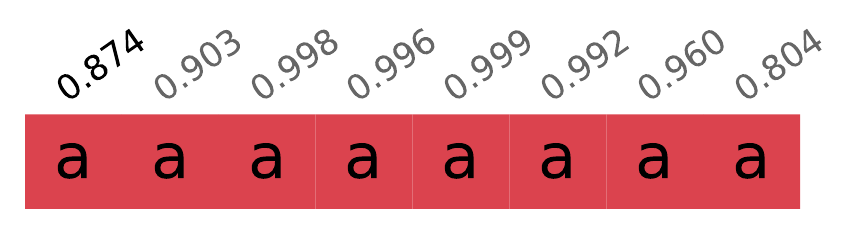}\end{subfigure}&
			\begin{subfigure}{0.4\columnwidth}\includegraphics[scale=0.4]{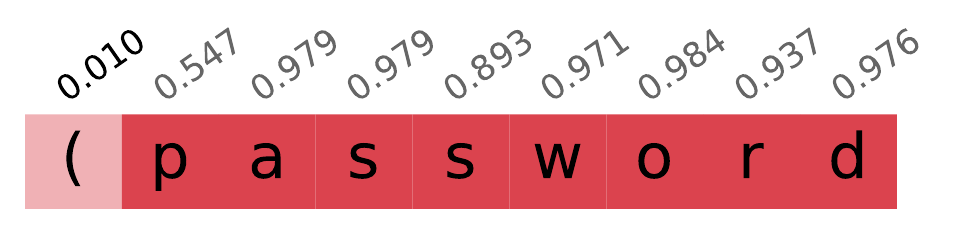}\end{subfigure}\\
			\begin{subfigure}{0.4\columnwidth}\includegraphics[scale=0.4]{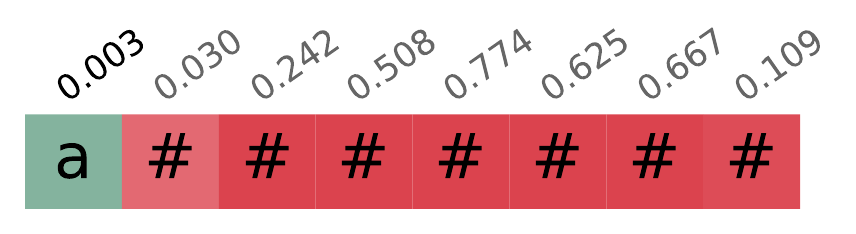}\caption{}\label{fig:meterpa}\end{subfigure}&
			\begin{subfigure}{0.4\columnwidth}\includegraphics[scale=0.4]{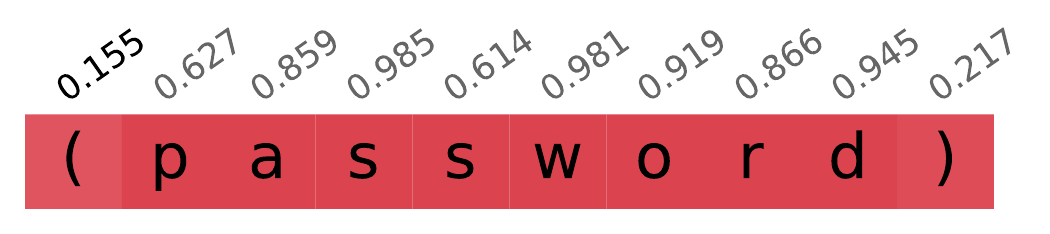}\caption{}\label{fig:meterpb}\end{subfigure}\\
		\end{tabular}
	}
	\caption{Estimated local conditional probabilities for two pairs of passwords. The numbers depicted above the strings report the $Q(x_i)$ value for each character (rounding applied).}
	\label{figure:meter_out_ex}
\end{figure}
\par
Eventually, such undirected formalization intrinsically weeds out all the limitations observed for the previous estimation process (\ie the Bayesian network in \reffig{fig:gma}). Now, every local measurement is computed within the context offered by any other symbol in the string. %Therefore, relations that were previously assessed as impossible can now be naturally described. %This statement becomes apparent as soon as we reconsider the examples made above in the discussion. 
In the example, $y\myeq \TT{aaaaaaa} ~/~ z \myeq \TT{a\scriptsize{\#\#\#\#\#\#}}$, indeed, the local conditional probability of the first character can be now backward-influenced from the context offered from the subsequent part of the string. This is clearly observable from the output of an instance of our meter reported in \reffig{fig:meterpa}, where the value of $Q(\x_1\myeq\T{a})$ drastically varies between the two cases, \ie $y$ and $z$. As expected, we have $Q(\x_1\myeq\T{a} | y) \gg Q(\x_1\myeq\T{a} | z)$ verified in the example. A similar intuitive result is reported in \reffig{fig:meterpb}, where the example $x\myeq\TT{(password)}$ is considered. Here, the meter first scores the string $x'\myeq\TT{(password }$, then it scores the complete password $x\myeq\TT{(password)}$. In this case, we expect that the presence of the last character $\T{)}$ would consistently influence the conditional measurement of the first bracket in the string. Such expectation is perfectly captured from the reported output, where appending at the end of the string the symbol~$\T{)}$ increases the probability of the first bracket of a factor $\sim 15$.
\par
However, obtaining these improvements does not come for free. Indeed, under the MRF construction, the productory over the local conditional probabilities (better defined as potential functions or factors within this context) does not provide the exact joint probability distribution of $\x$. Instead, such product results in a unnormalized version of it:
$P(\x) \propto \prod_{i=1}^{\ell} Q(\x_i)\myeq\tilde{P}(\x)$ with $P(x) = \frac{\tilde{P}(\x)}{Z}$. 
In the equation, $Z$ is the partition function. This result follows from the Hammersley–Clifford theorem \cite{hamme}.
Nevertheless, the unnormalized joint distribution preserves the core properties needed to the meter functionality. Most importantly, we have that: 
\begin{equation}
	\forall x, x'\ :\ P(x) \geq P(x') \Leftrightarrow \tilde{P}(x) \geq \tilde{P}(x') \quad .
\end{equation}
That is, if we sort a list of passwords according to the true joint $P(\x)$ or according to the unnormalized version $\tilde{P}(\x)$, we obtain the same identical ordering. Consequently, no deviation from the adversarial interpretation of PPSMs described in Section \ref{ppsm_intro} is implied. Indeed, we have $\ks_{P(\x)}=\ks_{\tilde{P}(\x)}$ for every password distribution, key-space, and suitable sorting function.
Furthermore, the joint probability distribution, if needed, can be approximated using suitable approximation methods, as discussed in Appendix \ref{app:est_gn}. Alternatively, we can use the energy-based models framework to capture the proposed procedure.\\

%Appendix \ref{sec:feedmech}, instead, reports a more detailed description of the feedback mechanism.\\
\new{It is important to highlight that, although we present our approach under the character-level description, our method can be directly applied to $n$-grams or tokens without any modification.}
\par
\subsubsection{Details on the password feedback mechanism and further applications}
\label{sec:feedmech}
Joint probability can be understood as a compatibility score assigned to a specific configuration of the MRF; it tells us the likelihood of observing
a sequence of characters during the interaction with the password generative process. On a smaller scale, a local conditional probability measures the impact that a single character has in the final security score. Namely, it indicates how much the character contributes to the probability of observing a certain password $x$. Within this interpretation, low-probabilities characters push the joint probability of $x$ to be closer to zero (secure), whereas high-probability characters (\ie $Q(x_1) \lesssim 1$) make no significant contribution to lowering the password probability (insecure). Therefore, users can strengthen their candidate passwords by substituting high-probability characters with suitable lower-probability ones (\eg Figure~\ref{fig:per_ex}).\\
%Henceforth, within this context, we refer to this process as password perturbation.\\
%In a general construction, during this process, the user can be left free to choose which symbol substitutes a high-probability character. However, while this choice eventually helps the user to compose a more usable password, it cannot guarantee the perturbation to convergence to a secure entry. 
\new{Unfortunately, users' perception of password security has been shown to be generally erroneous \cite{dousers}, and, without explicit guidelines, it would be difficult for them to select suitable lower-probability substitutes.} To address this limitation, one could adopt our approach based on local conditional distributions as an effective  mechanism to help users select  secure substitute symbols. Indeed, $\forall_i Q(\x_i)$ are able to clarify which symbol is a secure substitute and which is not for each character $x_i$ of $x$. In particular, a distribution $Q(\x_i)$, defined on the whole alphabet $\Sigma$, assigns a probability to every symbol $s$ that the character $\x_i$ can potentially assume. For a symbol $s\in\Sigma$, the probability $Q(\x_i\myeq s)$ measures how much the event $\x_i\myeq s$ is probable given all the observable characters in $x$. Under this interpretation, a candidate, secure substitution of $x_i$ is a symbol with very low $Q(\x_i\myeq s)$ (as this will lower the joint probability of $x$). In particular, every symbol $s$ s.t. $Q(\x_i\myeq s) < Q(\x_i\myeq x_i)$ given $x$ is a secure substitution for $x_i$.
Table~\ref{tab:guess_c} better depicts this intuition. The Table reports the alphabet sorted by $Q(\x_i)$ for each $x_i$ in the example password $x\myeq\TT{PaSsW0rD!}$. The bold symbols between parenthesis indicate $x_i$. Within this representation, all the symbols below the respective $x_i$ for each $\x_i$ are suitable substitutions that improve password strength. This intuition is empirically proven in Section~\ref{sec:exp_local_p}.
It is important to note that the suggestion mechanism must be randomized to avoid any bias in the final password distribution.\footnote{That is, if weak passwords are always perturbed in the same way, they will be easily guessed.} To this end, one can provide the user with $k$ random symbols among the pool of secure substitutions, \ie $\{s \mid Q(\x_i\myeq s) < Q(\x_i\myeq x_i)\}$.\\
\begin{table}[h!]
	\centering
	\caption{First seven entries of the ordering imposed on $\Sigma$ from the local conditional distribution for each character of the password $x\myeq\TT{PaSsW0rD!}$}
	\label{tab:guess_c}
	\vspace{1em}
	\resizebox{.95\columnwidth}{!}{%
		\begin{tabular}{c|c|c|c|c|c|c|c|c|c}
			\toprule
			{\scriptsize{Rank}} & \ppp{$\rc$aSsW0rD!} & \ppp{P$\rc$SsW0rD!} & \ppp{Pa$\rc$sW0rD!} & \ppp{PaS$\rc$W0rD!} & \ppp{PaSs$\rc$0rD!} & \ppp{PaSsW$\rc$rD!} & \ppp{PaSsW0$\rc$D!} & \ppp{PaSsW0r$\rc$!} & \ppp{PaSsW0rD$\rc$} \\
			\midrule
			\scriptsize{0} & \sem{P} & A & s & S & w & O & R & d & 1 \\
			\scriptsize{1} & S & \sem{a} & \sem{S} & \sem{s} & \sem{W} & o & \sem{r} & \sem{D} & S \\
			\scriptsize{2} & p & @ & c & A & \# & \sem{0} & N & t & 2 \\
			\scriptsize{3} & B & 3 & n & T & f & I & 0 & m & s \\
			\scriptsize{4} & C & 4 & t & E & k & i & L & l & 3 \\
			\scriptsize{5} & M & I & d & H & 1 & \# & D & k & \sem{!} \\
			\scriptsize{6} & 1 & 1 & r & O & F & A & n & e & 5 \\
			7 & c & 5 & x & \$ & 3 & @ & X & r & 9 \\
			8 & s & 0 & \$ & I & 0 & ) & S & f & 4 \\
			$\vdots$ & $\vdots$ & $\vdots$ & $\vdots$ & $\vdots$ & $\vdots$ & $\vdots$ & $\vdots$ & $\vdots$ & $\vdots$\\
			%9 & F & O & i & Z & l & 3 & W & n & 6 \\
			\bottomrule
		\end{tabular}
	}
\end{table}
%A more detailed analysis of the probabilistic description of our meter is reported in section \ref{app:prob_disc} in Appendix.
%
%Finally, based on such measurements of probability, we can construct our feedback mechanism. An elegant interpretation of the latter can be found under the energy-based models perspective \cite{EBM}. Here, the model assigns minimal energies to the common configurations (\ie common passwords) and high energies to uncommon ones. To strengthen the chosen password, the user has to perturb the MRF configuration to increase its energy (\ie lowering the probability of the password). In doing so, the user is aware of the contribution that every character has in the system's global energy. Therefore, he/she can efficiently reduce the energy of the configuration by perturbing low-energy characters first. A more detailed analysis of the feedback mechanism is reported in \ref{sec:feedmech}.\\
 
In summary, in this section, we presented and motivated an estimation process able to unravel the feedback mechanism described in Section \ref{section:feedback_def}. Maintaining a purely theoretical focus, no information about the implementation of such methodology has been offered to the reader. Next, in Section \ref{section:dp_impl}, we describe how such a meter can be shaped via an efficient deep learning framework. 
\vspace{-5pt}

\section{Meter implementation}
\label{section:dp_impl}
In this section, we present a deep-learning-based implementation of the estimation process introduced in Section \ref{section:mrf_def}. Here, we describe the model and its training process. Then, we explain how the trained network can be used as a building block for the proposed password meter.
\vspace{-5pt}
\paragraph{\textbf{Model training.}}
From the discussion in Section \ref{section:mrf_def}, our procedure requires the parametrization of an exponentially large number of interactions among random variables. Thus, any tabular approach, such as the one used from Markov Chains or PCFG \cite{PCFG}, is \textit{a priori} excluded for any real-world case. To make such a meter feasible, we reformulate the underlying estimation process so that it can be approximated with a neural network. In our approach, we simulate the Markov Random Field described in Section \ref{section:mrf_def} using a deep convolutional neural network trained to compute $Q(\x_i)$ (Eq. \ref{eq:q}) for each possible configuration of the structured model. 
In doing so, we train our network to solve an \textit{inpainting}-like task defined over the textual domain. Broadly speaking, inpainting is the task of reconstructing missing information from mangled inputs, mostly images with missing or damaged patches \cite{inpainting}. 
%A good inpainting model must be able to infer missing content leveraging the context maintained from the observable data. 
\textbf{Under the probabilistic perspective, the model is asked to return a probability distribution over all the unobserved elements of $x$, explicitly measuring the conditional probability of those concerning the observable context}. 
%Therefore, consequently, performing a good approximation of the data probability distribution describing the underlying domain. 
Therefore, the network has to disentangle and model the semantic relation among all the factors describing the data (\eg characters in a string) to reconstruct input instances correctly.\\
Generally, the architecture and the training process used for inpainting tasks resemble an auto-encoding structure \cite{AE}. 
%That is, an autoencoder network \cite{AE} is trained to learn a form of reconstruction function over the interested domain. 
In the general case, these models are trained to revert self-induced damage carried out on instances of a train-set $\ds$. At each training step, an instance $x\in \ds$ is artificially mangled with an information-destructive transformation to create a mangled variation $\tilde{x}$. Then, the network, receiving $\tilde{x}$ as input, is optimized to produce an output that most resembles the original $x$; that is, the network is trained to reconstruct~$x$~from~$\tilde{x}$.\\
In our approach, we train a network to infer missing characters in a mangled password. In particular, we iterate over a password leak (\ie our train-set) by creating mangled passwords and train the network to recover them. The mangling operation is performed by removing a randomly selected character from the string. For example, the train-set entry $x\myeq\TT{iloveyou}$ is transformed in $\tilde{x}\myeq$``ilov$\rc$you" if the $5$'th character is selected for deletion, where the symbol \T{$\rc$} represents the \textit{``empty character"}. A compatible proxy-task has been previously used in \cite{IPGVRL} to learn a suitable password representation for guessing attacks.\\
\input{inference_scheme}
We chose to model our network with a deep \textit{residual} structure arranged to create an autoencoder. The network follows the same general Context Encoder \cite{CAE} architecture defined in \cite{IPGVRL} with some modifications. 
%The encoder and the decoder are composed of the concatenation of the same number of deep residual bottleneck blocks \cite{resnet}. T
To create an information bottleneck, the encoder connects with the decoder through a latent space junction obtained through two fully connected layers. We observed that enforcing a latent space, and a prior on that, consistently increases the meter effectiveness. For that reason, we maintained the same regularization proposed in \cite{IPGVRL}; a maximum mean discrepancy regularization that forces a standard normal distributed latent space. The final loss function of our model is reported in \refeq{eq:ae_loss}. In the equation, $Enc$ and $Dec$ refer to the encoder and decoder network respectively, $s$ is the \textit{softmax} function applied row-wise\footnote{The Decoder outputs $\ell$ estimations; one for each input character. Therefore, we apply the softmax function separately on each of those to create $\ell$ probability distributions.}, the distance function $d$ is the cross-entropy, and $mmd$ refers to the \textit{maximum mean discrepancy}.
\begin{equation}
\label{eq:ae_loss}
\mathbb{E}_{x,\tilde{x}}[d(x,\ s(Dec(Enc(\tilde{x})))] + \alpha \mathbb{E}_{z\sim N(0, \mathbb{I})}[mmd(z, Enc(\tilde{x}))]
\end{equation}
Henceforth, we refer to the composition of the encoder and the decoder as $f(x)=s(Dec(Enc(x)))$. %Given its autoencoder nature, $f$ produces as outputs $\ell$ probability distributions over the alphabet $\Sigma$ \ie one for each character in the input string\footnote{Distributions imposed from the softmax function (see \refeq{eq:ae_loss}).}. However, at inference time, we are interested only in the $i$'th distribution; the one attributed to the removed character $x_i$ \ie $Q(\x_i \mid \tilde{x})$
We train the model on the widely adopted \textit{RockYou} leak \cite{rockyou} considering an $80/20$ train-test split. From it, we filter passwords presenting fewer than $5$ characters. We train different networks considering different maximum password lengths, namely, $16$, $20$, and $30$. In our experiments, we report results obtained with the model trained on a maximum length equal to $16$, as no substantial performance variation has been observed among the different networks.
Eventually, we produce three neural nets with different architectures; a large network requiring $36\text{MB}$ of disk space, a medium-size model requiring $18\text{MB}$, and a smaller version of the second that requires $6.6\text{MB}$. These models can be potentially further compressed using the same quantization and compression techniques harnessed in \cite{FLA}.\footnote{The code, pre-trained models, and other materials related to our work are publicly available at: \url{https://github.com/pasquini-dario/InterpretablePPSM}.}
%Fine-grained information about the used architectures and hyper-parameters are reported in Appendix \ref{app:archhyper}. We implement our approach using the \textit{TensorFlow} framework. All the experiments have been carried out on a \textit{Nvidia DGX-2} machine. 
\vspace{-5pt}
\paragraph{\textbf{Model inference process.}}
Once the model is trained, we can use it to compute the conditional probability $Q(x_i)$ (\refeq{eq:q}) for each $i$ and each possible configuration of the MRF. This is done by querying the network $f$ using the same mangling trick performed during the training. The procedure used to compute $Q(x_i)$ for $x$ is summarized in the following steps:
\vspace{-2pt}
\begin{enumerate}
	\itemsep0em 
	\item We substitute the $i$'th character of $x$ with the \textit{empty character} '$\rc$', obtaining a mangled password $\tilde{x}$. 
	\item Then, we feed $\tilde{x}$ to a network that outputs a probability distribution over $\Sigma$ of the unobserved random variable $\x_i$ \ie $Q(\x_i)$.
	\item Given $Q(\x_i)$, we marginalize out $x_i$, obtaining the probability:\\ $Q(x_i)~\myeq~P(\x_i\myeq x_i~\mid~\tilde{x})$.
\end{enumerate}
\vspace{-3pt}
For instance, if we want to compute the local conditional probability of the character $\T{e}$ in the password $x=\TT{iloveyou}$, we first create $\tilde{x}=$``ilov$\rc$you" and use it as input for the net, obtaining $Q(\x_5)$, then we marginalize that (\ie $Q(\x_5 \myeq \T{e})$) getting the probability $P(\x_5 \myeq \T{e} \mid \tilde{x})$. From the probabilistic point of view, this process is equivalent to fixing the observable variables in the MRF and querying the model for an estimation of the single unobserved character.

At this point, to cast both the feedback mechanism defined in Section~\ref{section:feedback_def} and the unnormalized joint probability of the string, we have to measure $Q(x_i)$ for each character $x_i$ of the tested password. This is easily achieved by repeating the inference operation described above for each character comprising the input string. A graphical representation of this process is depicted in \reffig{fig:inference}. It is important to highlight that the $\ell$ required inferences are independent, and their evaluation can be performed in parallel (\ie batch level parallelism), introducing almost negligible overhead over the single inference. Additionally, with the use of a \mbox{feed-forward network}, we  avoid the sequential computation that is intrinsic in recurrent networks (\eg the issue afflicting \cite{FLA}), and that can be excessive for a reactive client-side implementation. Furthermore, the convolutional structure enables the construction of very deep neural nets with a limited memory footprint.\\
\vspace{-5pt}

In conclusion, leveraging the trained neural network, we can compute the potential of each factor/vertex in the Markov Random Field (defined as local conditional probabilities in our construction). As a consequence, we are now able to cast a PPSM featuring the character-level feedback mechanism discussed in Section \ref{section:feedback_def}. 
%Next, in Section \ref{section:dp_impl}, we briefly cover the feedback mechanism. 
Finally, in Section \ref{section:results}, we empirically evaluate the soundness of the proposed meter. 
\vspace{-5pt}
\section{Evaluation}
\label{section:results}
\input{result}
\vspace{-5pt}
\paragraph{\textbf{Limitations.}}\new{Since the goal of our evaluation was mainly to validate the soundness of the proposed estimation process, we did not perform user studies and we did not evaluate human-related factors such as password memorability although we recognize their importance.} 
\vspace{-8pt}
\section{Conclusion}
\vspace{-5pt}
%\todo{add limitations}
\label{sec:conclusion}

In this paper, we showed that it is possible to construct interpretable probabilistic password meters by  rethinking the underlying password mass estimation. We presented an undirected probabilistic interpretation of the password generative process that can be used to build precise and sound password feedback mechanisms. Moreover, we demonstrated that such an estimation process could be instantiated via a lightweight deep learning implementation. We validated our undirected description and deep learning solution by showing that our meter achieves comparable accuracy with other existing approaches while introducing a unique character-level feedback mechanism that generalizes any  heuristic construction. 

%\balance
\bibliographystyle{plain}
\bibliography{bib.bib}

\input{appendix}
\end{document}

%% file: intro.tex
Accurately measuring password strength is essential to guarantee the security of password-based authentication systems. Even more critical, however, is training users to select secure passwords in the first place. 
One common approach is to rely on password policies that list a series of requirements for a strong password. This approach is limited or even harmful \cite{passwordExhaustion}. Alternatively, Passwords Strength Meters (PSMs) have been shown to be useful and are witnessing increasing adoption in commercial solutions \cite{measure_up, ontheaccuracy}.\\
The first instantiations of PSMs were based on simple heuristic constructions. Password strength was estimated via either handcrafted features such as \textit{LUDS} (which counts lower and uppercase letters, digits, and symbols) or heuristic entropy definitions. Unavoidably, given their heuristic nature, this class of PSMs failed to accurately measure password security \cite{empirical, testing}.\\
More recently, thanks to an active academic interest, PSMs based on more sound constructions and rigorous security definitions have been proposed. In the last decade, indeed, a considerable research effort gave rise to more precise meters capable of accurately measuring password strength \cite{FLA, fuzzyPSM, MM}.\\
However, meters have also become proportionally more opaque and inherently hard to interpret due to the increasing complexity of the employed approaches.
State-of-the-art solutions base their estimates on blackbox parametric probabilistic models \cite{FLA, MM} that leave no room for interpretation of the evaluated passwords; they do not provide any feedback to users on what is wrong with their password or how to improve it. 
We advocate for explainable approaches in password meters, where users receive additional insights and become cognizant of which parts of their passwords could straightforwardly improve. This makes the password selection process less painful since users can keep their passwords of choice mostly unchanged while ensuring they are secure.\\
\vspace{-5pt}

In the present work, we show that the same rigorous probabilistic framework capable of accurately measuring password strength can also fundamentally describe the relation between password security and password structure. By rethinking the underlying mass estimation process, we create the first {\em interpretable probabilistic password strength} meter. Here, the password probability measured by our meter can be decomposed and used to estimate further the strength of every single character of the password.
This explainable approach allows us to assign a security score to each atomic component of the password and determine its contribution to the overall security strength. This evaluation is, in turn, returned to the user who can tweak a few "weak" characters and consistently improve the password strength against guessing attacks. Figure~\ref{fig:per_ex} illustrates the selection process.
In devising the proposed mass estimation process, we found it ideally suited for being implemented via a deep learning architecture. In the paper, we show how that can be cast as an efficient client-side meter employing deep convolutional neural networks. Our work's major contributions are: (i) We formulate a novel password probability estimation framework based on undirected probabilistic models. (ii) We show that such a framework can be used to build a precise and sound password feedback mechanism. (iii) We implement the proposed meter via an efficient and lightweight deep learning framework ideally suited for client-side operability.

%% file: related.tex
Here, we briefly review early approaches to the definition of PSMs. We focus on the most influential works as well as to the ones most related to ours.
\vspace{-5pt}
\paragraph{Probabilistic PSMs:}
 Originally thought for guessing attacks \cite{mm_first}, Markov model approaches have found natural application in the password strength estimation context.
 Castelluccia~\etal~\cite{MM} use a stationary, finite-state Markov chain as a direct password mass estimator. Their model computes the joint probability by separately measuring the conditional probability of each pair of $n$-grams in the observed passwords.
 Melicher~\etal~\cite{FLA} extended the Markov model approach by leveraging a character/token level Recurrent Neural Network (RNN) for modeling the probability of passwords. 
 As discussed in the introduction, pure probabilistic approaches are not capable of any natural form of feedback. In order to partially cope with this shortcoming, a hybrid approach has been investigated in \cite{FLA2}. Here, the model of Melicher~\etal~\cite{FLA} is aggregated with a series of $21$ heuristic, hand-crafted feedback mechanisms such as detection of \textit{leeting behaviors} or common tokens (\eg keyboard walks).\\
 Even if harnessing a consistently different form of feedback, our framework merges these solutions into a single and jointly learned model. Additionally, in contrast with \cite{FLA2}, our feedback has a concrete probabilistic interpretation as well as complete freedom from any form of human bias. Interestingly enough, our model autonomously learns some of the heuristics hardwired in \cite{FLA2}. For instance, our model learned that capitalizing characters in the middle of the string could consistently improve password strength.
\par
 \paragraph{Token look-up PSMs:}
 Another relevant class of meters is that based on the token look-up approach. Generally speaking, these are non-parametric solutions that base their strength estimation on collections of sorted lists of tokens like leaked passwords and word dictionaries. Here, a password is modeled as a combination of tokens, and the relative security score is derived from the ranking of the tokens in the known dictionaries. Unlike probabilistic solutions, token-based PSMs are able to return feedback to the user, such as an explanation for the weakness of a password relying on the semantic attributed to the tokens composing the password.
% Such feedback is mainly based on the semantic attributed to the tokens composing the password.
A leading member of token look-up meters is \textit{zxcvbn} \cite{zxcvbn}, which assumes a password as a combination of tokens such as \textit{token, reversed, sequence repeat, keyboard, and date}.
This meter scores passwords according to a heuristic characterization of the guess-number \cite{gue_and_en}. Such score is described as the number of combinations of tokens necessary to match the tested password by traversing the sorted tokens lists.\\
\textit{zxcvbn} is capable of feedback. For instance, if one of the password components is identified as \textit{``repeat}", \textit{zxcvbn} will recommend the user to avoid the use of repeated characters in the password. Naturally, this kind of feedback mechanism inherently lacks generality and addresses just a few human-chosen scenarios. 
%\textit{zxcvbn} is available through a lightweight implementation.  
As discussed by the authors themselves, \textit{zxcvbn} suffers from various limitations. By assumption, it is unable to model the relationships among different patterns occurring in the same passwords. Additionally, like other token look-up based approaches, it fails to coherently model unobserved patterns and tokens.\\
Another example of token look-up approach is the one proposed in \cite{telepathwords}. \textit{Telepathwords} discourages a user from choosing weak passwords by predicting the next most probable characters during the password typing. In particular, predicted characters are shown to the user in order to dissuade him/her from choosing them as the next characters in the password. These are reported together with an explanation of why those characters were predicted. However, as for \textit{zxcvbn}, such feedback solely relies on hardwired scenarios (for instance, the use of profanity in the password). \textit{Telepathwords} is server-side only.

%% file: gm.tex
\begin{figure}[t]
	\centering
	\resizebox{.81\columnwidth}{!}{%
		\begin{subfigure}[t]{0.6\columnwidth}
			\centering
			\begin{tikzpicture}[->,>=stealth',shorten >=1pt,auto,node distance=1.2cm,
			thick,main node/.style={circle,draw,font=\sffamily\Large\bfseries}]
			
			\node[main node] (1) {$x_1$};
			\node[main node] (2) [below  of=1] {$x_2$};
			\node[main node] (3) [below right of=2] {$x_3$};
			\node[main node] (4) [ right of=3] {$x_4$};
			
			\path[every node/.style={font=\sffamily\small}]
			(1) edge [right] node[left] {} (2)
			(2) edge node {} (3)
			(3) edge node {} (4);
			%(1) edge[bend left] node [left] {} (4)
			%(1) edge[bend left] node [left] {} (3)
			%(2) edge[bend right=90] node [right] {} (4);
			\end{tikzpicture}
			\caption{}
			\label{fig:gma}
		\end{subfigure}
		\begin{subfigure}[t]{0.6\columnwidth}
			\centering
			\begin{tikzpicture}[->,>=stealth',shorten >=1pt,auto,node distance=1.5cm,
			thick,main node/.style={circle,draw,font=\sffamily\Large\bfseries}]
			
			\node[main node] (1) {$x_1$};
			\node[main node] (2) [below left of=1] {$x_2$};
			\node[main node] (3) [below right of=2] {$x_3$};
			\node[main node] (4) [below right of=1] {$x_4$};
			
			\path[-]
			(1) edge node [left] {} (4)
			edge [right] node[left] {} (2)
			edge [left] node {} (3)
			(2) edge node {} (4)
			(3) edge node {} (4)
			(3) edge node {} (2);
			\end{tikzpicture}	
			\caption{}
			\label{fig:gmb}
		\end{subfigure}
	}
\caption{Two graphical models describing different interpretations of the generative probability distribution for passwords of length four. Graph (a) represents a Bayesian network. Scheme (b) depicts a Markov Random Field.}
\label{figure:gm}
 \end{figure}

%% file: inference_scheme.tex
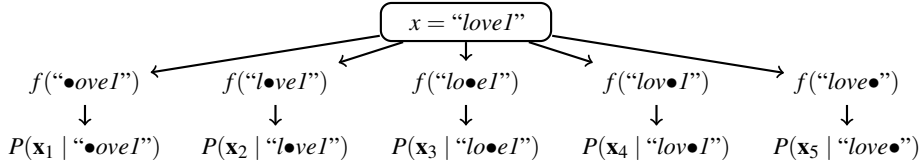
\begin{figure*}[t]
	\centering
	\resizebox{.7\textwidth}{!}{%
\begin{tikzpicture}[thick,scale=1]
	\tikzstyle{every node}=[font=\small]
	\tikzstyle{n} = [rectangle, minimum width=1cm, minimum height=.5cm,text centered]
	\tikzstyle{root} = [rectangle,
	rounded corners,
	draw=black,  thick,
	text width=2cm,
	minimum height=.5cm,
	text centered]
	%[rectangle, minimum width=1cm, minimum height=1cm,text centered]
	
	\tikzstyle{arrow} = [->,
	line width=0.8pt,
	shorten <=0pt,
	shorten >=0pt]
	%[thick,->,>=stealth]
	
	%\draw[step=1cm,gray,very thin] (-8,-4) grid (8,4);
		
	\node (start) [root, yshift=6cm] {$x=\TT{love1}$};
	
	\node (x0) [n, below of=start, xshift=-5cm, yshift=0.2cm] {$f(\TT{\rcc ove1})$};
	\node (p0) [n, below of=x0, yshift=.15cm] {$P(\x_1 \mid \TT{\rcc ove1})$};
	
	\node (x1) [n, right of=x0, xshift=1.5cm] {$f(\TT{l\rcc ve1})$};
	\node (p1) [n, below of=x1, yshift=0.15cm] {$P(\x_2 \mid \TT{l\rcc ve1})$};
	
	\node (x2) [n, right of=x1, xshift=1.5cm] {$f(\TT{lo\rcc e1})$};
	\node (p2) [n, below of=x2, yshift=.15cm] {$P(\x_3 \mid \TT{lo\rcc e1})$};
	
	\node (x3) [n, right of=x2, xshift=1.5cm] {$f(\TT{lov\rcc1})$};
	\node (p3) [n, below of=x3, yshift=.15cm] {$P(\x_4  \mid \TT{lov\rcc1})$};
	
	\node (x4) [n, right of=x3, xshift=1.5cm] {$f(\TT{love\rcc})$};
	\node (p4) [n, below of=x4, yshift=.15cm] {$P(\x_5  \mid \TT{love\rcc})$};

	\draw [arrow] (start) -- (x0);
	\draw [arrow] (x0) -- (p0);
	\draw [arrow] (start) -- (x1);
	\draw [arrow] (x1) -- (p1);
	\draw [arrow] (start) -- (x2);
	\draw [arrow] (x2) -- (p2);
	\draw [arrow] (start) -- (x3);
	\draw [arrow] (x3) -- (p3);
	\draw [arrow] (start) -- (x4);
	\draw [arrow] (x4) -- (p4);
\end{tikzpicture}
}
\caption{Graphical depiction of the complete inference process for the password $x\myeq\TT{love1}$. The function $f$ refers to the trained autoencoder and the symbol~'$\rc$' refers to the deleted character.}
\label{fig:inference}
\end{figure*}

%% file: result.tex
In this section, we empirically validate the proposed estimation process as well as its deep learning implementation. First, in Section \ref{sec:accuracy}, we evaluate the capability of the meter of accurately assessing password strength at string-level.
% In the process, we compare with \sota PPSMs, showing our meter capable of 
Next, in Section \ref{sec:exp_local_p}, we demonstrate the intrinsic ability of the local conditional probabilities of being sound descriptors of password strength at character-level.
\vspace{-5pt}
\subsection{Measuring meter accuracy}
\vspace{-5pt}
\label{sec:accuracy}
In this section, we evaluate the accuracy of the proposed meter at estimating password probabilities. To that purpose, following the adversarial reasoning introduced in Section \ref{ppsm_intro}, we compare the password ordering derived from the meter with the one from the ground-truth password distribution. In doing so, we rely on the guidelines defined in \cite{ontheaccuracy} for our evaluation. In particular, given a test-set (\ie a password leak), we consider a weighted rank correlation coefficient between ground-truth ordering and that derived from the meter. 
%The latter is obtained by applying the meter on each password of the test-set and sorting those according to the (unnormalized) joint probability. 
The ground-truth ordering is obtained by sorting the unique entries of the test-set according to the frequency of the password observed in the leak. In the process, we compare our solution with other fully probabilistic meters.
%, namely, Markov Models and the Neural approach reported in \cite{FLA}. 
A detailed description of the evaluation process follows.
\vspace{-5pt}
\paragraph{\textbf{Test-set.}} For modeling the ground-truth password distribution, we rely on the password leak discovered by 4iQ in the Dark Web\cite{BC_leak} on 5th December 2017. It consists of the aggregation of $\sim250$ leaks, consisting of $1.4$ billion passwords in total.
%, including well-known entries such as Linkedin, Myspace, and RockYou and novel breaches. In total, the set counts $1.4$ billion compressed passwords. 
In the cleaning process, we collect passwords with length in the interval $5-16$, obtaining a set of $\sim4 \cdot 10^8$ unique passwords that we sort in decreasing frequency order. Following the approach of \cite{ontheaccuracy}, we filter out all the passwords with a frequency lower than $10$ from the test-set.
%, as rare passwords could bring to erroneous measurement. 
Finally, we obtain a test-set composed of $10^{7}$ unique passwords that we refer to as $\BC$. Given both the large number of entries and the heterogeneity of sources composing it, we consider $\BC$ an accurate description of real-world passwords distribution.
\vspace{-5pt}
\paragraph{\textbf{Tested Meters.}} In the evaluation process, we compare our approach with other probabilistic meters. In particular:
\begin{itemize}
\itemsep0em 
\item The Markov model \cite{omen} implemented in \cite{nemo_git} (the same used in \cite{ontheaccuracy}). We investigate different $n$-grams configurations, namely, $2$-grams, $3$-grams and $4$-grams that we refer to as $\text{MM}_2$, $\text{MM}_3$ and $\text{MM}_4$, respectively. For their training, we employ the same train-set used for our meter. %Eventually, we obtain three models $\text{MM}_2$, $\text{MM}_3$ and $\text{MM}_4$ requiring $1.1\text{MB}$, $94\text{MB}$ and $8.8\text{GB}$ of disk space respectively.
\item The neural approach of Melicher~\etal~\cite{FLA}. We use the implementation available at \cite{FLA_git} to train the main architecture advocated in \cite{FLA}, \ie an RNN composed of three LSTM layers of $1000$ cells each, and two fully connected layers. 
The training is carried out on the same train-set used for our meter. We refer to the model as FLA.
% Eventually, we obtain a model composing of $60\text{MB}$ of parameters that we refer to as FLA. 
\end{itemize}
%\vspace{-5pt}
\paragraph{\textbf{Metrics.}} We follow the guidelines defined by Golla and {Dürmuth} \cite{ontheaccuracy} for evaluating the meters. We use the \textbf{weighted} \textit{Spearman} correlation coefficient (ws) to measure the accuracy of the orderings produced by the tested meters, as this has been demonstrated to be the most reliable correlation metric within this context \cite{ontheaccuracy}. This metric is defined as
\[
\text{ws}(t, m) = \frac{\sum_i^n[w_i(t_i-\bar{t})(m_i-\bar{m})]}{\sqrt{\sum_i^n[w_i(t-\bar{t}_i)^2] \sum_i^n [w_i(m-\bar{m}_i)^2]}} \quad ,
\]
where $t$ and $m$ are the sequence of rank assigned to the test-set from the ground-truth distribution and the tested meter, respectively, and where the bar notation (\eg $\bar{t}$) expresses the weighted mean in consideration of the sequence of weights~$w$. The weights are computed as the normalized inverse of the \mbox{ground-truth} ranks (\refeq{eq:w}).
\begin{equation}
	\label{eq:w}
	w = \frac{q}{\sum_i^n q_i} \quad \text{with} \quad q = \frac{1}{t + 1} \quad .
\end{equation}
In this metric, the weighting increases the relevance of weak passwords (\ie the ones with small ranks) in the score computation; that is, the erroneous placing of weak passwords (\ie asserting a weak password as strong) is highly penalized.
Unlike \cite{ontheaccuracy}, given the large cardinality and diversity of this leak, we directly use the ranking derived from the password frequencies in $\BC$ as ground-truth. Here, passwords with the same frequency value have received the same rank in the computation of the correlation metric.
\vspace{-7pt}
\paragraph{\textbf{Results.}} 
	Table~\ref{tab:ws} reports the measured correlation coefficient for each tested meter. In the table, we also report the required storage as auxiliary metric.\\
	Our meters, even the smallest, achieve higher or comparable score than the most performant Markov Model, \ie $\text{MM}_4$. On the other hand, our largest model cannot directly exceed the accuracy of the \sota estimator FLA, obtaining only comparable results. However, FLA requires more disk space than ours. Indeed, interestingly, our convolutional implementation permits the creation of remarkably lightweight meters. As a matter of fact, our smallest network shows a comparable result with $\text{MM}_4$ requiring more than a magnitude less disk space.\\
	In conclusion, the results confirm that the probability estimation process defined in Section \ref{section:mrf_def} is indeed sound and capable of accurately assessing password mass at string-level. The proposed meter shows comparable effectiveness with the \sota \cite{FLA}, whereas, in the \textit{large} setup, it outperforms standard approaches such as Markov Chains. Nevertheless, we believe that even more accurate estimation can be achieved by investigating deeper architectures and/or by performing hyper-parameters tuning over the model.
\begin{table}[t!]
	\centering
				\caption{Rank correlation coefficient computed between $\BC$ and the tested meters.}
				\label{tab:ws}
				\resizebox{1\columnwidth}{!}{%
					\begin{tabular}{r||cccc|ccc}
					&\textbf{ $\text{MM}_2$ }& \textbf{$\text{MM}_3$} & \textbf{$\text{MM}_4$} & \textbf{FLA }& \makecell{\textbf{ours}\\(large)} & \makecell{\textbf{ours}\\(middle)} & \makecell{\textbf{ours}\\(small)} \\\toprule
					\textbf{ \makecell{Weighted\\Spearman $\uparrow$}} & 0.154 & 0.170 & 0.193 & 0.217 & 0.207 & 0.203 & 0.199\\ \midrule
					\textbf{\makecell{Required \\Disk Space $\downarrow$} } & 1.1MB & 94MB & 8.8GB & 60MB & 36MB & 18MB & 6.6MB \\
					\bottomrule
				\end{tabular}
			}
	\end{table}
\vspace{-10pt}
\subsection{Analysis of the relation between local conditional probabilities and password strength}
\vspace{-5pt}
\label{sec:exp_local_p}
In this section, we test the capability of the proposed meter to correctly model the relation between password structure and password strength. In particular, we investigate the ability of the measured local conditional probabilities of determining the tested passwords' insecure components.\\
Our evaluation procedure follows three main steps. Starting from a set of weak passwords $X$:
\begin{enumerate}
\item We perform a guessing attack on $X$ in order to estimate the guess-number of each entry of the set.
\item For each password $x \in X$, we substitute $n$ characters of $x$ according to the estimated local conditional probabilities (\ie we substitute the characters with highest $Q(\x_i)$), producing a perturbed password $\tilde{x}$.
\item We repeat the guessing attack on the set of perturbed passwords and measure the variation in the attributed guess-numbers.
\end{enumerate}

Hereafter, we provide a detailed description of the evaluation procedure.
\vspace{-5pt}
\paragraph{\textbf{Passwords sets.}} The evaluation is carried out considering a set of weak passwords. In particular, we consider the first $10 ^{4}$ most frequent passwords of the $\BC$ set.
\vspace{-5pt}
\paragraph{\textbf{Password perturbations.}}
In the evaluation, we consider three types of password perturbation:\\ \textbf{(1)} The first acts as a baseline and consists of the substitution of random positioned characters in the passwords with randomly selected symbols. Such a general strategy is used in \cite{FLA2} and \cite{persuasion} to improve the user's password at composition time. The perturbation is applied by randomly selecting $n$ characters from $x$ and substituting them with symbols sampled from a predefined character pool. \new{In our simulations, the pool consists of the $25$ most frequent symbols in $\BC$ (\ie mainly lowercase letters and digits). Forcing this character-pool aims at preventing the tested perturbation procedures to create artificially complex passwords such as strings containing extremely uncommon \textit{unicode} symbols.} We refer to this perturbation procedure as \textbf{Baseline}.\\
\textbf{(2)} The second perturbation partially leverages the local conditional probabilities induced by our meter. Given a password $x$, we compute the conditional probability $Q(x_i)$ for each character in the string. Then, we select and substitute the character with maximum probability, \ie $\argmax_{x_i} Q(x_i)$. The symbol we use in the substitution is randomly selected from the same pool used for the baseline perturbation (\ie top-$25$ frequent symbols). When $n$ is greater than one, the procedure is repeated sequentially using the perturbed password obtained from the previous iteration as input for the next step. 
%Examples of modified passwords are reported in the appendix.
We refer to this procedure as \textbf{Semi-Meter}.\\
\textbf{(3)} The third perturbation extends the second one by exploiting the local conditional distributions. Here, as in the Semi-Meter-based, we substitute the character in $x$ with the highest probability. However, rather than choosing a substitute symbol in the pool at random, we select that according to the distribution $Q(\x_i)$, where $i$ is the position of the character to be substituted. In particular, we choose the symbol the minimize $Q(\x_i)$, \ie $\argmin_{s \in \Sigma'} Q(\x_i\myeq s)$, where $\Sigma'$ is the allowed pool of symbols. We refer to this method as \textbf{Fully-Meter}. %Examples of perturbed passwords are reported in the Appendix.
\begin{table}[t!]
	\centering
	\caption{Strength improvement induced by different perturbations.  The last two rows of the table report the AGI ratio between the two meter-based approaches and the baseline.
	}
	\label{table:presult}
	\resizebox{.85\columnwidth}{!}{%
		\begin{tabular}{r|c|c|cc}
			%& \multicolumn{3}{c||}{ $\BC^1$}       \\ 
			& $n = 1$  & $n=2$   & $n=3$   \\ \toprule
			Baseline (PNP)  & 0.022 & 0.351 & 0.549 & \\  
			Semi-Meter (PNP) & 0.036 & 0.501 & 0.674 \\
			Fully-Meter (PNP)  & 0.066 & 0.755 & 0.884 \\ \midrule
			Baseline \textbf{(AGI)}  & $3.0\cdot 10^{10}$ & $3.6\cdot 10^{11}$ & $5.6\cdot 10^{11}$ \\
			Semi-Meter \textbf{(AGI)} & $4.6\cdot 10^{10}$ & $5.1\cdot 10^{11}$ & $6.8\cdot 10^{11}$ \\
			Fully-Meter \textbf{(AGI)}  & $8.2\cdot 10^{10}$ & $7.7\cdot 10^{11}$ & $8.9\cdot 10^{11}$ \\ \midrule
			Semi-Meter / Baseline \textbf{(AGI)} & \textbf{1.530} & \textbf{1.413} & \textbf{1.222} \\
			Fully-Meter / Baseline \textbf{(AGI)} & \textbf{2.768} & \textbf{2.110} & \textbf{1.588} \\ \bottomrule  
		\end{tabular}
	}
\end{table}
\vspace{-5pt}
\paragraph{\textbf{Guessing Attack.}} We evaluate password strength using the \textit{min-auto} strategy advocated in \cite{min-guess}. 
%The ensemble we employ is composed of HashCat \cite{hashcat}, PCFG \cite{PCFG, PCFG_git} and the Markov chain approach implemented in \cite{omen, omen_git}. We limit each tool to produce $10^{10}$ guesses. The total size of the generated guesses is $\sim 3$TB. 
 Here, guessing attacks are simultaneously performed with different guessing tools, and the guess-number of a password is considered the minimum among the attributed guess-numbers. In performing such attacks, we rely on the combination of three widely adopted solutions, namely, HashCat \cite{hashcat}, PCFG \cite{PCFG, PCFG_git} and the Markov chain approach proposed in \cite{omen, omen_git}. For tools requiring a training phase, \ie OMEN and PCFG, we use the same train-set used for our model (\ie 80\% of \textit{RockYou}). Similarly, for HashCat, we use the same data set as input dictionary\footnote{In this case, passwords are unique and sorted in decreasing frequency.} and \textit{generated2} as rules set. During the guesses generation, we maintain the default settings of each implementation. We limit each tool to produce $10^{10}$ guesses. The total size of the generated guesses is $\sim 3$TB.
 \vspace{-5pt}
\paragraph{\textbf{Metrics.}} In the evaluation, we are interested in measuring the increment of password strength caused by an applied perturbation. We estimate that value by considering the Average Guess-number Increment (henceforth, referred to as~AGI); that is, the average delta between the guess-number of the original password and the guess-number of the perturbed password: \[\text{AGI}(X)~\myeq~\frac{1}{|X|} \sum_{i=0}^{|X|}[ g(\tilde{x}^i)~-~g(x^i)]\]
where $g$ is the guess-number, and $\tilde{x}^i$ refers to the perturbed version of the $i$'th password in the test set. During the computation of the guess-numbers, it is possible that we fail to guess a password. In such a case, we attribute an artificial guess-number equals to $10^{12}$ to the un-guessed passwords. Additionally, we consider the average number of un-guessed passwords as an ancillary metrics; we refer to it with the name of Percentage Non-Guessed Passwords~(PNP) and~compute~it~as:
\[\text{PNP(X)}~\myeq~\frac{1}{|X|} |\{x^i \mid g(x^i) \neq \bot \ \land \ g(\tilde{x}^i)=\bot \}|,\]
where $g(x) = \bot$ when $x$ is not guessed during the guessing attack.
\vspace{-5pt}
\paragraph{\textbf{Results.}} We perform the tests over three values of $n$ (\ie the number of perturbed characters), namely, $1$, $2$, and $3$. Results are summarized in Table~\ref{table:presult}. The AGI caused by the two meter-based solutions is always greater than that produced by random perturbations. On average, that is twice more effective with respect to the Fully-Meter baseline and about $35\%$ greater for the Semi-Meter. 
The largest relative benefit is observable when $n=1$, \ie a single character is modified. Focusing on the Fully-Meter approach, indeed, the guidance of the local conditional probabilities permits a guess-number increment $~2.7$ times bigger than the one caused by a random substitution in the string. This advantage drops to $\sim1.5$ when $n=3$, since, after two perturbations, passwords tend to be already out of the dense zone of the distribution. Indeed, at $n=3$ about $88\%$ of the passwords perturbed with the Fully-Meter approach cannot be guessed during the guessing attack (\ie PNP). This value is only $\sim55\%$ for the baseline. More interestingly, the results tell us that substituting two ($n=2$) characters following the guide of the local conditional probabilities causes a guess-number increment greater than the one obtained from three ($n=3$) random perturbations. As a matter of fact, the AGI for the Fully-Meter perturbation is $\sim7.6\cdot 10 ^{11}$ for $n=2$ whereas is $\sim5.7\cdot 10^{11}$ for the baseline when $n=3$.\\
In the end, these results confirm that the local conditional distributions are indeed sound descriptors of password security at the structural level.% Following the guide of these values, can move towards the creation of secure passwords using limit effort.

%% file: appendix.tex
\section*{Appendix}
\setcounter{section}{0}
%\counterwithin{table}{section}
% or try \arabic{section}
%\counterwithin{figure}{section}
\renewcommand{\thesection}{\Alph{section}}%
%reset equation to appendix number
%\setcounter{equation}{0}
%\renewcommand{\theequation}{\Alph{section}.\arabic{equation}}
%
\section{Estimating guess-numbers}
\label{app:est_gn}
Within the context of PPSMs, a common solution to approximate guess-numbers \cite{gue_and_en} is using the Monte Carlo method proposed in \cite{montecarlo_g}. With a few adjustments, the same approach can be applied to our meter. In particular, we have to derive an approximation of the partition function $Z$. This can be done by leveraging the Monte Carlo method as follows:
\begin{equation}
\label{eq:mm_z}
Z \simeq N \cdot \mathbb{E}_{x}[P(x)]
\end{equation}
where $N$ is the number of possible configurations of the MRF (\ie the cardinality of the key-space), and $x$ is a sample from the posterior distribution of the model. Samples from the model can be obtained in three ways: (1) sampling from the latent space of the autoencoder (as done in \cite{IPGVRL}), (2) performing Gibbs sampling from the autoencoder, or (3) using a dataset of passwords that follow the same distribution of the model.
Once we have an approximation of $Z$, we can use it to normalize every joint probability, \ie $P(x) = \frac{\tilde{P}(x)}{Z}$ and then apply the method in \cite{montecarlo_g}. Alternatively, we could adopt a more articulate solution as in \cite{pmlr-v31-ma13a}. 
In any event, the estimation of the partition function $Z$ is performed only once and can be done offline.
\section{Model Architectures and hyper-parameters}
\label{app:archhyper}
\begin{figure}
	\centering
	\includegraphics[scale=.33]{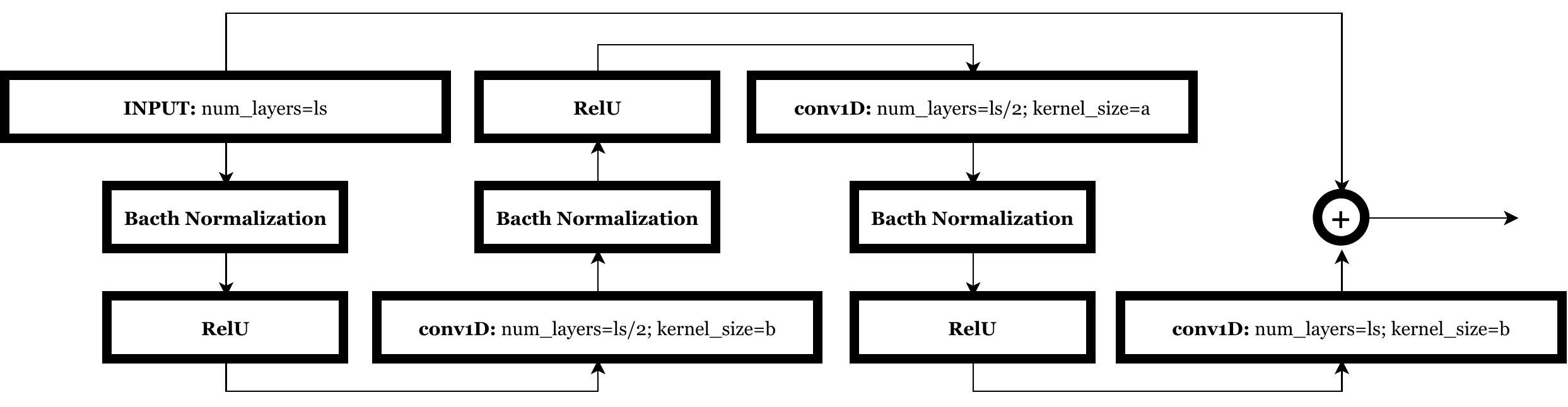}
	\caption{\small Depiction of: \rb{ls}{(a, b)} }
	\label{fig:resblock}
\end{figure}

In this section, we detail the technical aspects of our deep learning implementation.\\
\textbf{Architectures}
As previously described, we base our networks on a \textit{resnet} structure. We use a bottleneck residual block composed of three mono-dimensional convolutional layers as the atomic building block of the networks. A graphical description of that is depicted in Figure~\ref{fig:resblock}. We construct three different networks with different sizes (intended as the number of trainable parameters). We determine the size of the networks by varying the number of residual blocks, the kernel size of the convolutional layers in the blocks, and the number of filters. The three architectures are reported in Tables~\ref{table:cwaearcha}, \ref{table:cwaearchb} and \ref{table:cwaearchc}.\\
\textbf{Training process}
Table~\ref{table:hyperparametercwae} reports the used hyper-parameters. During the training, we apply label smoothing, which is controlled from the parameter $\epsilon$. We found our models taking particular advantage from large batch-sizes. We limit that to $3072$ for technical limitations; however, we believe that bigger batches could further increment the quality of the password estimation.
\begin{table}
	\centering
	\caption{Hyper-parameters used to train our AE}%\vspace{3mm}
	\resizebox{.5\columnwidth}{!}{%
		\begin{tabular}{lc}
			\toprule
			\multicolumn{1}{l}{\textbf{Hyper-parameter}} & \multicolumn{1}{l}{\textbf{Value}} \\ \midrule
			$\alpha$ & 10 \\  
			Batch size & 3024 \\ 
			Learning rate & 0.0001 \\ 
			Optimizer & \textit{Adam} \\ 		
			Train Epochs & \makecell{$\text{small}\myeq10$ \\ $\text{medium}\myeq5$ \\ $\text{large}\myeq5$}\\ 
			$\epsilon$ & 0.05 \\
			\bottomrule
		\end{tabular}
	}
	\label{table:hyperparametercwae}
\end{table}
\begin{table}[h]
	\centering
	\caption{Small architecture.}
	\resizebox{.85\columnwidth}{!}{%
		\begin{tabular}{ll}
		 \toprule
			\cd{3}{128}{same}{linear} \\ 
			\rb{128}{(3, 1)}  \\ 
			\rb{128}{(3, 1)}  \\ 
			\rb{128}{(3, 1)}  \\ 
			\rb{128}{(3, 1)}  \\ 
			\rb{128}{(3, 1)}  \\ 
			\rb{128}{(3, 1)}  \\ 
			\texttt{Flatten} \\
			\texttt{FullyConnected}[128, \textit{linear}] \\ 
			\texttt{FullyConnected}[$\textit{MaxPasswordLength}\cdot128$, \textit{linear}]  \\ 
			\texttt{Reshape}[\textit{MaxPasswordLength}, 128]  \\ 
			\rb{128}{(3, 1)}  \\ 
			\rb{128}{(3, 1)}  \\ 
			\rb{128}{(3, 1)}  \\ 
			\rb{128}{(3, 1)}  \\ 
			\rb{128}{(3, 1)}  \\ 
			\rb{128}{(3, 1)}  \\ 
			\texttt{Flatten} \\
			\texttt{FullyConnected}[$\textit{MaxPasswordLength}\cdot\textit{AlphabetCardinality}$, \textit{linear}]  \\ \toprule
					\end{tabular}
		}
	\label{table:cwaearcha}
\end{table}

			\begin{table}[h]
				\centering
				\caption{Medium architecture.}
				\resizebox{.85\columnwidth}{!}{%
					\begin{tabular}{ll}
			 \toprule
			\cd{5}{128}{same}{linear} \\ 
			\rb{128}{(5, 3)}  \\ 
			\rb{128}{(5, 3)}  \\  
			\rb{128}{(5, 3)}  \\ 
			\rb{128}{(5, 3)}  \\ 
			\rb{128}{(5, 3)}  \\ 
			\rb{128}{(5, 3)}  \\ 
			\texttt{Flatten} \\
			\texttt{FullyConnected}[80, \textit{linear}] \\ 
			\texttt{FullyConnected}[$\textit{MaxPasswordLength}\cdot128$, \textit{linear}]  \\ 
			\texttt{Reshape}[\textit{MaxPasswordLength}, 128]  \\ 
			\rb{128}{(5, 3)}  \\ 
			\rb{128}{(5, 3)}  \\  
			\rb{128}{(5, 3)}  \\ 
			\rb{128}{(5, 3)}  \\ 
			\rb{128}{(5, 3)}  \\ 
			\rb{128}{(5, 3)}  \\
			\texttt{Flatten} \\
			\texttt{FullyConnected}[$\textit{MaxPasswordLength}\cdot\textit{AlphabetCardinality}$, \textit{linear}]  \\ \toprule
			
					\end{tabular}
		}
	\label{table:cwaearchb}
\end{table}
\begin{table}[h!]
	\centering
	\caption{Large architecture.}
	\resizebox{.85\columnwidth}{!}{%
		\begin{tabular}{ll}
			 \toprule
			\cd{5}{128}{same}{linear} \\ 
			\rb{200}{(5, 3)}  \\ 
			\rb{200}{(5, 3)}  \\  
			\rb{200}{(5, 3)}  \\ 
			\rb{200}{(5, 3)}  \\ 
			\rb{200}{(5, 3)}  \\ 
			\rb{200}{(5, 3)}  \\  
			\rb{200}{(5, 3)}  \\ 
			\rb{200}{(5, 3)}  \\ 
			\texttt{Flatten} \\
			\texttt{FullyConnected}[80, \textit{linear}] \\ 
			\texttt{FullyConnected}[$\textit{MaxPasswordLength}\cdot128$, \textit{linear}]  \\ 
			\texttt{Reshape}[\textit{MaxPasswordLength}, 128]  \\ 
			\rb{200}{(5, 3)}  \\ 
			\rb{200}{(5, 3)}  \\  
			\rb{200}{(5, 3)}  \\ 
			\rb{200}{(5, 3)}  \\ 
			\rb{200}{(5, 3)}  \\ 
			\rb{200}{(5, 3)}  \\  
			\rb{200}{(5, 3)}  \\ 
			\rb{200}{(5, 3)}  \
			\texttt{Flatten} \\
			\texttt{FullyConnected}[$\textit{MaxPasswordLength}\cdot\textit{AlphabetCardinality}$, \textit{linear}]  \\ \toprule
			
		\end{tabular}
	}
	\label{table:cwaearchc}
\end{table}
\section{Supplementary resources}
\begin{figure}[h]
	\centering
	\caption{Additional examples of the feedback mechanism.}
	\label{fig:fmex}
	\resizebox{1\columnwidth}{!}{%
		\begin{tabular}{cc}
\begin{subfigure}{1\columnwidth}\centering\includegraphics[scale=0.45]{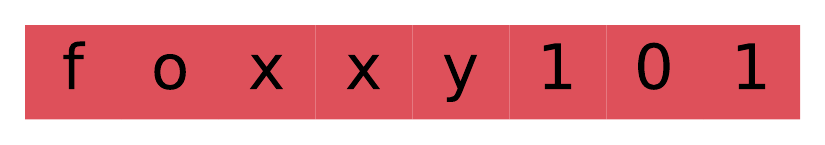}\end{subfigure}\\
\begin{subfigure}{1\columnwidth}\centering\includegraphics[scale=0.45]{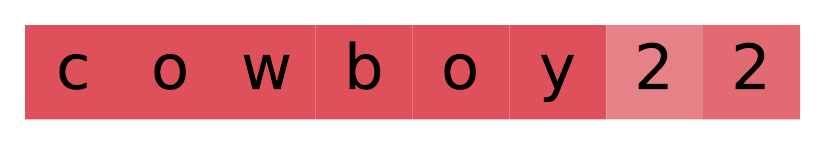}\end{subfigure}\\
\begin{subfigure}{1\columnwidth}\centering\includegraphics[scale=0.45]{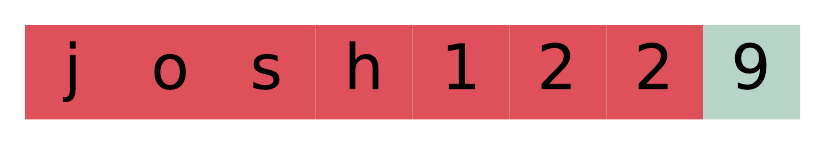}\end{subfigure}\\
\begin{subfigure}{1\columnwidth}\centering\includegraphics[scale=0.45]{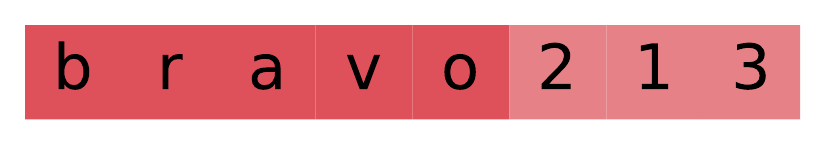}\end{subfigure}\\
\begin{subfigure}{1\columnwidth}\centering\includegraphics[scale=0.45]{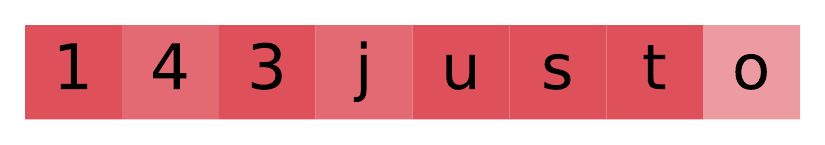}\end{subfigure}\\
\begin{subfigure}{1\columnwidth}\centering\includegraphics[scale=0.45]{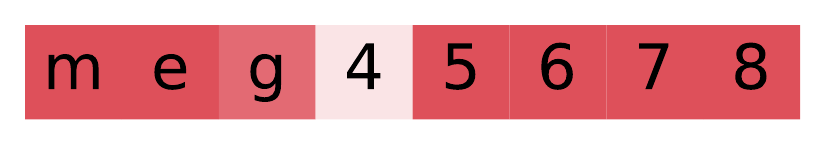}\end{subfigure}\\
\begin{subfigure}{1\columnwidth}\centering\includegraphics[scale=0.45]{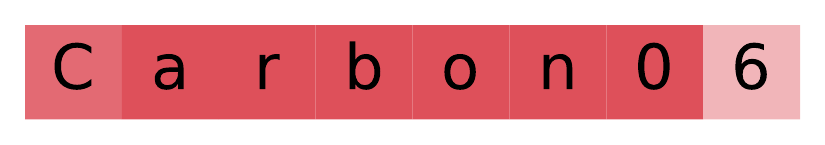}\end{subfigure}\\
\begin{subfigure}{1\columnwidth}\centering\includegraphics[scale=0.45]{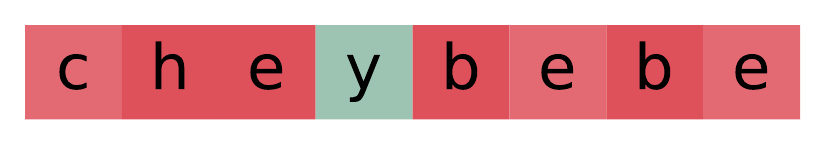}\end{subfigure}\\
\begin{subfigure}{1\columnwidth}\centering\includegraphics[scale=0.45]{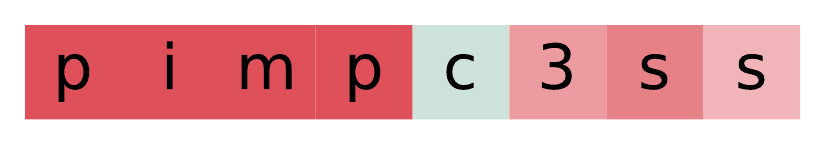}\end{subfigure}\\
\begin{subfigure}{1\columnwidth}\centering\includegraphics[scale=0.45]{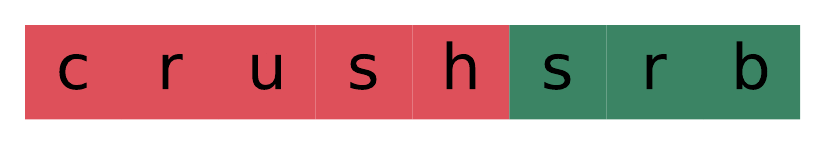}\end{subfigure}\\
\begin{subfigure}{1\columnwidth}\centering\includegraphics[scale=0.45]{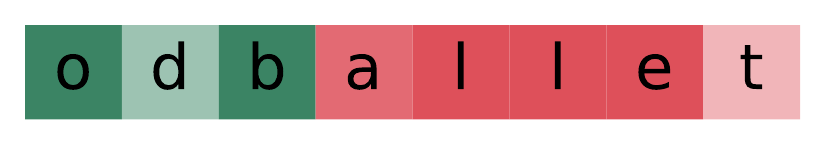}\end{subfigure}\\
\begin{subfigure}{1\columnwidth}\centering\includegraphics[scale=0.45]{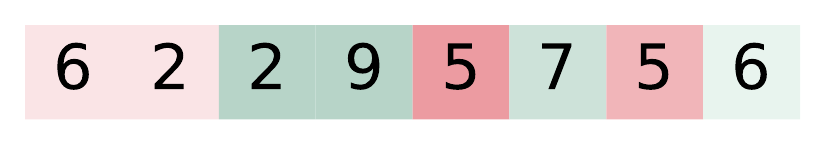}\end{subfigure}\\
\begin{subfigure}{1\columnwidth}\centering\includegraphics[scale=0.45]{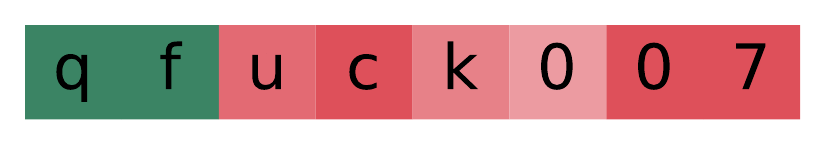}\end{subfigure}\\
\begin{subfigure}{1\columnwidth}\centering\includegraphics[scale=0.45]{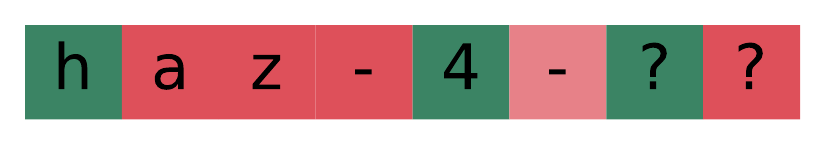}\end{subfigure}\\
\begin{subfigure}{1\columnwidth}\centering\includegraphics[scale=0.45]{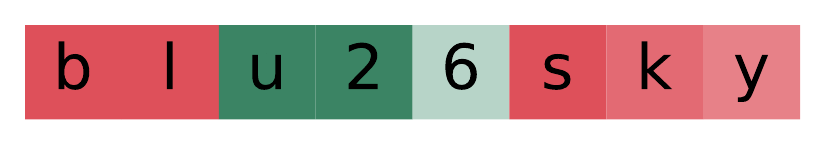}\end{subfigure}\\
\begin{subfigure}{1\columnwidth}\centering\includegraphics[scale=0.45]{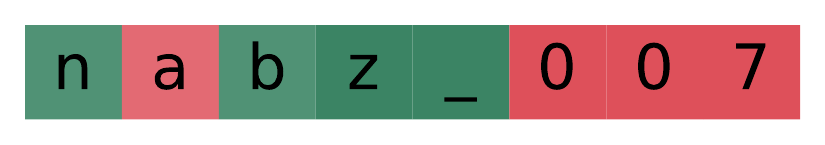}\end{subfigure}\\
\begin{subfigure}{1\columnwidth}\centering\includegraphics[scale=0.45]{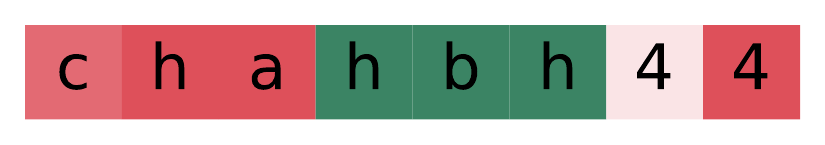}\end{subfigure}\\
\begin{subfigure}{1\columnwidth}\centering\includegraphics[scale=0.45]{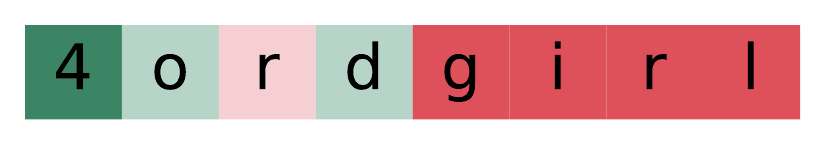}\end{subfigure}\\
\begin{subfigure}{1\columnwidth}\centering\includegraphics[scale=0.45]{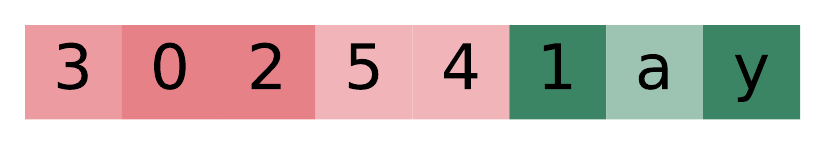}\end{subfigure}\\
\begin{subfigure}{1\columnwidth}\centering\includegraphics[scale=0.45]{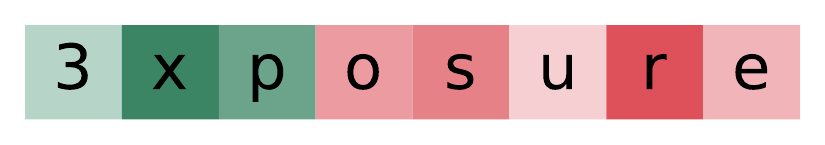}\end{subfigure}\\
\begin{subfigure}{1\columnwidth}\centering\includegraphics[scale=0.45]{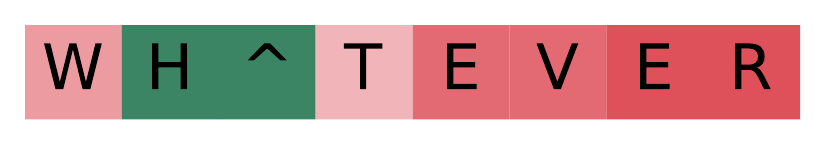}\end{subfigure}\\
\begin{subfigure}{1\columnwidth}\centering\includegraphics[scale=0.45]{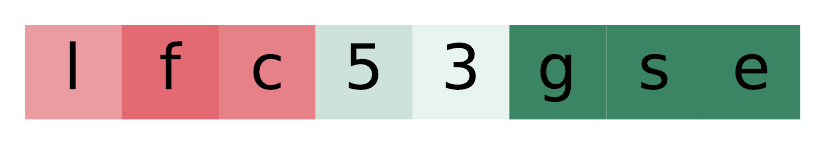}\end{subfigure}\\
\begin{subfigure}{1\columnwidth}\centering\includegraphics[scale=0.45]{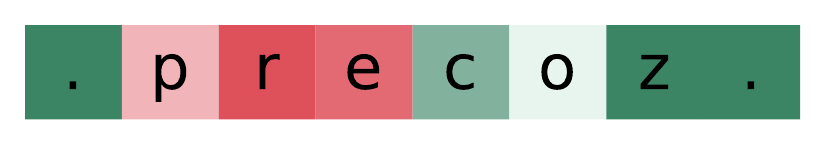}\end{subfigure}\\
\begin{subfigure}{1\columnwidth}\centering\includegraphics[scale=0.45]{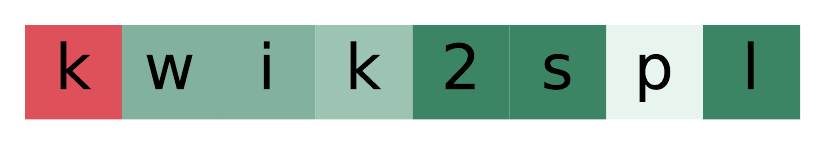}\end{subfigure}\\
\begin{subfigure}{1\columnwidth}\centering\includegraphics[scale=0.45]{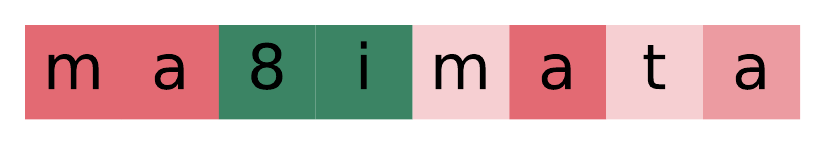}\end{subfigure}\\
\begin{subfigure}{1\columnwidth}\centering\includegraphics[scale=0.45]{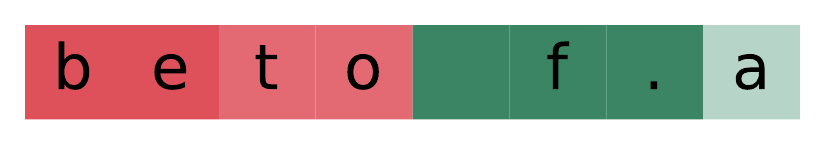}\end{subfigure}\\
\begin{subfigure}{1\columnwidth}\centering\includegraphics[scale=0.45]{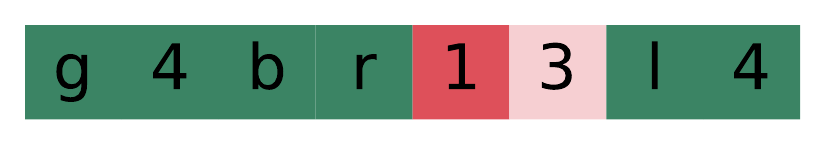}\end{subfigure}\\
\begin{subfigure}{1\columnwidth}\centering\includegraphics[scale=0.45]{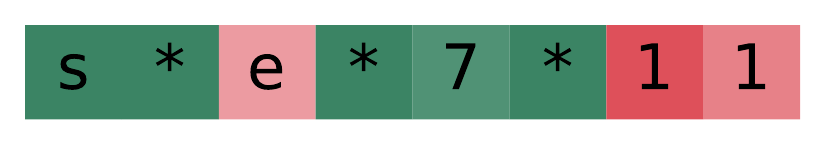}\end{subfigure}\\
\begin{subfigure}{1\columnwidth}\centering\includegraphics[scale=0.45]{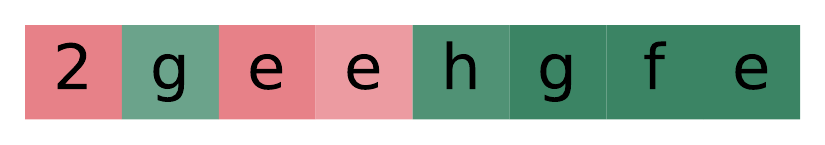}\end{subfigure}\\
\begin{subfigure}{1\columnwidth}\centering\includegraphics[scale=0.45]{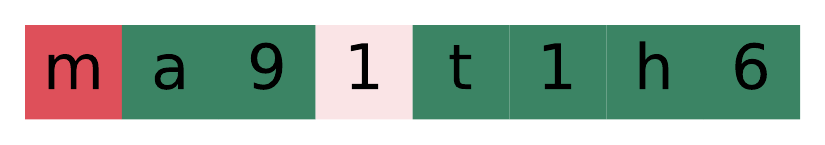}\end{subfigure}\\
\begin{subfigure}{1\columnwidth}\centering\includegraphics[scale=0.45]{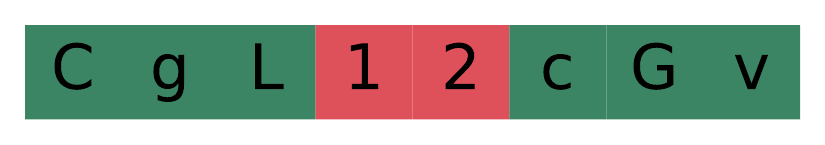}\end{subfigure}\\
\begin{subfigure}{1\columnwidth}\centering\includegraphics[scale=0.45]{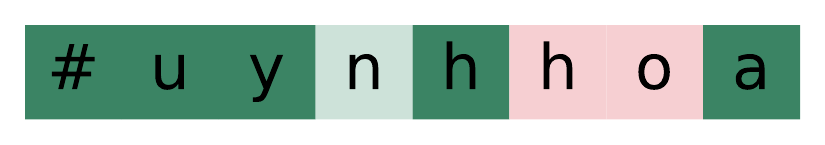}\end{subfigure}\\
		\end{tabular}
	}
\end{figure}

This Section reports additional resources. 
	Table~\ref{tab:per_acc} reports examples of password perturbation performed using the method \textbf{Fully-meter} on the three values of $n$. The example passwords (first column) are sampled from $\BC$. Figure~\ref{fig:fmex} reports additional examples of the feedback mechanism. The depicted passwords have been randomly sampled from the tail of the \textit{RockYou} leak.
\begin{table}
	\caption{Examples of password perturbation \textbf{automatically} produced by the method \textbf{Fully-meter}. Symbols in bold are the ones substituted by the meter.}
	\label{tab:per_acc}
\begin{tabular}{c||c|c|c}
	\toprule
	$x$ &       $n\myeq1$ &           $n\myeq2$ &                $n\myeq3$ \\
	\midrule
	heaven7 &   \textbf{2}eaven7 &   \textbf{2}eav\textbf{1}n7 &   \textbf{2}e\textbf{9}v\textbf{1}n7 \\
	corvette &   corv\textbf{l}tte &   c\textbf{5}rv\textbf{l}tte &   c\textbf{5}rv\textbf{l}tt\textbf{b} \\
	mariah &    m\textbf{3}riah &    m\textbf{3}\textbf{u}iah &    \textbf{u}\textbf{3}\textbf{u}iah \\
	373737 &    37373\textbf{l} &    \textbf{t}7373\textbf{l} &    \textbf{t}737\textbf{u}\textbf{l} \\
	veronica1 &  v\textbf{s}ronica1 &  v\textbf{s}ron\textbf{5}ca1 &  v\textbf{s}r\textbf{s}n\textbf{5}ca1 \\
	ariana1 &   aria\textbf{o}a1 &   \textbf{3}ria\textbf{o}a1 &   \textbf{3}r\textbf{6}a\textbf{o}a1 \\
	goodgirl &   goodgir\textbf{3} &   g\textbf{9}odgir\textbf{3} &   \textbf{u}\textbf{9}odgir\textbf{3} \\
	cheer &    c\textbf{9}eer &    c\textbf{9}e\textbf{h}r &    c\textbf{9}\textbf{y}\textbf{h}r \\
	mahalko &   m\textbf{l}halko &   m\textbf{l}h\textbf{8}lko &   m\textbf{l}h\textbf{8}lk\textbf{t} \\
	19981998 &   19981\textbf{n}98 &   \textbf{u}9981\textbf{n}98 &   \textbf{u}9\textbf{o}81\textbf{n}98 \\
	123456aa &   \textbf{i}23456aa &   \textbf{i}234\textbf{r}6aa &   \textbf{i}23\textbf{n}\textbf{r}6aa \\
	helena &    hele\textbf{a}a &    h\textbf{4}le\textbf{a}a &    h\textbf{4}\textbf{i}e\textbf{a}a \\
	montana1 &   monta\textbf{2}a1 &   mo\textbf{3}ta\textbf{2}a1 &   mo\textbf{3}\textbf{a}a\textbf{2}a1 \\
	vancouver &  vancouv\textbf{m}r &  va\textbf{9}couv\textbf{m}r &  va\textbf{9}co\textbf{6}v\textbf{m}r \\
	fuck12 &    f\textbf{h}ck12 &    f\textbf{h}ck1\textbf{n} &    f\textbf{h}c\textbf{8}1\textbf{n} \\
	patriots1 &  pat\textbf{9}iots1 &  pat\textbf{9}iot\textbf{i}1 &  p\textbf{5}t\textbf{9}iot\textbf{i}1 \\
	evelyn1 &   eve\textbf{4}yn1 &   \textbf{6}ve\textbf{4}yn1 &   \textbf{6}ve\textbf{4}yn\textbf{r} \\
	pancho &    panc\textbf{2}o &    pa\textbf{6}c\textbf{2}o &    \textbf{9}a\textbf{6}c\textbf{2}o \\
	malibu &    m\textbf{5}libu &    m\textbf{5}\textbf{y}ibu &    m\textbf{5}\textbf{y}ib\textbf{6} \\
	ilovemysel &  ilo\textbf{0}emysel &  i\textbf{i}o\textbf{0}emysel &  i\textbf{i}o\textbf{0}emys\textbf{8}l \\
	galatasaray & galatasar\textbf{4}y & galat\textbf{6}sar\textbf{4}y & g\textbf{8}lat\textbf{6}sar\textbf{4}y \\
	tootsie1 &   to\textbf{5}tsie1 &   to\textbf{5}ts\textbf{9}e1 &   to\textbf{5}t\textbf{n}\textbf{9}e1 \\
	sayangku &   saya\textbf{8}gku &   s\textbf{3}ya\textbf{8}gku &   s\textbf{3}ya\textbf{8}gk\textbf{8} \\
	moneyman &   mo\textbf{5}eyman &   mo\textbf{5}eym\textbf{d}n &   \textbf{u}o\textbf{5}eym\textbf{d}n \\
	theboss &   th\textbf{9}boss &   th\textbf{9}bos\textbf{4} &   th\textbf{9}b\textbf{t}s\textbf{4} \\
	112211 &    \textbf{o}12211 &    \textbf{o}1221\textbf{u} &    \textbf{o}\textbf{e}221\textbf{u} \\
	k12345 &    k123\textbf{y}5 &    k12\textbf{o}\textbf{y}5 &    k\textbf{n}2\textbf{o}\textbf{y}5 \\
	alexis &    \textbf{9}lexis &    \textbf{9}l\textbf{r}xis &    \textbf{9}l\textbf{r}x\textbf{o}s \\
	princess7 &  princ\textbf{4}ss7 &  pri\textbf{h}c\textbf{4}ss7 &  pr\textbf{r}\textbf{h}c\textbf{4}ss7 \\
	rooster1 &   roo\textbf{3}ter1 &   roo\textbf{3}t\textbf{l}r1 &   r\textbf{6}o\textbf{3}t\textbf{l}r1 \\
	june15 &    jun\textbf{m}15 &    jun\textbf{m}\textbf{r}5 &    j\textbf{l}n\textbf{m}\textbf{r}5 \\
	samurai1 &   \textbf{0}amurai1 &   \textbf{0}amu\textbf{0}ai1 &   \textbf{0}\textbf{e}mu\textbf{0}ai1 \\
	surfer1 &   s\textbf{9}rfer1 &   s\textbf{9}rf\textbf{n}r1 &   s\textbf{9}rf\textbf{n}r\textbf{3} \\
	lokomotiv &  l\textbf{h}komotiv &  l\textbf{h}komot\textbf{6}v &  l\textbf{h}ko\textbf{8}ot\textbf{6}v \\
	rfn.irf &   rfn.i\textbf{5}f &   \textbf{5}fn.i\textbf{5}f &   \textbf{5}\textbf{e}n.i\textbf{5}f \\
	melisa &    m\textbf{t}lisa &    m\textbf{t}lis\textbf{l} &    m\textbf{t}l\textbf{6}s\textbf{l} \\
	minime &    \textbf{3}inime &    \textbf{3}inim\textbf{t} &    \textbf{3}i\textbf{i}im\textbf{t} \\
	peaceout &   pea\textbf{a}eout &   \textbf{8}ea\textbf{a}eout &   \textbf{8}ea\textbf{a}eo\textbf{1}t \\
	louise &    lo\textbf{4}ise &    l\textbf{r}\textbf{4}ise &    l\textbf{r}\textbf{4}is\textbf{r} \\
	Liverpool &  Live\textbf{h}pool &  Live\textbf{h}p\textbf{2}ol &  Li\textbf{6}e\textbf{h}p\textbf{2}ol \\
	147896 &    1\textbf{d}7896 &    1\textbf{d}78\textbf{y}6 &    1\textbf{d}78\textbf{y}\textbf{y} \\
	aditya &    ad\textbf{l}tya &    \textbf{4}d\textbf{l}tya &    \textbf{4}d\textbf{l}ty\textbf{i} \\
	qwerty13 &   qw\textbf{m}rty13 &   qw\textbf{m}r\textbf{9}y13 &   qw\textbf{m}r\textbf{9}y\textbf{u}3 \\
	070809 &    \textbf{i}70809 &    \textbf{i}708\textbf{d}9 &    \textbf{i}7\textbf{r}8\textbf{d}9 \\
	emmanuel1 &  emm\textbf{9}nuel1 &  emm\textbf{9}nue\textbf{i}1 &  e\textbf{0}m\textbf{9}nue\textbf{i}1 \\
	beautiful2 &  be\textbf{1}utiful2 &  be\textbf{1}utif\textbf{1}l2 &  be\textbf{1}ut\textbf{n}f\textbf{1}l2 \\
			\iffalse	
	123456789h &  12\textbf{b}456789h &  1\textbf{h}\textbf{b}456789h &  1\textbf{h}\textbf{b}45678\textbf{4}h \\
	jeffhardy1 &  je\textbf{6}fhardy1 &  je\textbf{6}fh\textbf{5}rdy1 &  je\textbf{6}fh\textbf{5}rdy\textbf{u} \\
	madness &   madn\textbf{6}ss &   m\textbf{8}dn\textbf{6}ss &   m\textbf{8}dn\textbf{6}s\textbf{l} \\
	diamonds1 &  d\textbf{t}amonds1 &  d\textbf{t}amonds\textbf{l} &  d\textbf{t}amo\textbf{9}ds\textbf{l} \\
\fi
	\bottomrule
\end{tabular}
\end{table}

%% file: main.bbl
\begin{thebibliography}{10}

\bibitem{BC_leak}
{``1.4 Billion Clear Text Credentials Discovered in a Single Database''}.
\newblock \url{https://tinyurl.com/t8jp5h7}.

\bibitem{hashcat}
{``hashcat GitHub''}.
\newblock \url{https://github.com/hashcat}.

\bibitem{nemo_git}
{``NEMO Markov Model GitHub''}.
\newblock \url{https://github.com/RUB-SysSec/NEMO}.

\bibitem{FLA_git}
{``Neural Cracking GitHub''}.
\newblock \url{https://github.com/cupslab/neural\_network\_cracking}.

\bibitem{omen_git}
{``OMEN GitHub''}.
\newblock \url{https://github.com/RUB-SysSec/OMEN}.

\bibitem{PCFG_git}
{``PCFG GitHub''}.
\newblock \url{https://github.com/lakiw/pcfg\_cracker}.

\bibitem{rockyou}
{``RockYou Leak''}.
\newblock \url{https://downloads.skullsecurity.org/passwords/rockyou.txt.bz2}.

\bibitem{AE}
Pierre Baldi.
\newblock Autoencoders, unsupervised learning, and deep architectures.
\newblock In {\em Proceedings of ICML Workshop on Unsupervised and Transfer
  Learning}, volume~27 of {\em Proceedings of Machine Learning Research}, pages
  37--49, Bellevue, Washington, USA, 02 Jul 2012. PMLR.

\bibitem{MM}
Claude Castelluccia, Markus D{\"u}rmuth, and Daniele Perito.
\newblock {Adaptive Password-Strength Meters from Markov Models}.
\newblock In {\em NDSS}, 2012.

\bibitem{passwordExhaustion}
Luke~St. Clair, Lisa Johansen, William Enck, Matthew Pirretti, Patrick Traynor,
  Patrick McDaniel, and Trent Jaeger.
\newblock Password exhaustion: Predicting the end of password usefulness.
\newblock In {\em Information Systems Security}, pages 37--55, Berlin,
  Heidelberg, 2006. Springer Berlin Heidelberg.

\bibitem{empirical}
M.~{Dell' Amico}, P.~{Michiardi}, and Y.~{Roudier}.
\newblock Password strength: An empirical analysis.
\newblock In {\em 2010 Proceedings IEEE INFOCOM}, pages 1--9, March 2010.

\bibitem{montecarlo_g}
Matteo Dell’Amico and Maurizio Filippone.
\newblock {Monte Carlo Strength Evaluation: Fast and Reliable Password
  Checking}.
\newblock In {\em Proceedings of the 22nd ACM SIGSAC Conference on Computer and
  Communications Security}, CCS ’15, page 158–169, New York, NY, USA, 2015.
  Association for Computing Machinery.

\bibitem{omen}
Markus Duermuth, Fabian Angelstorf, Claude Castelluccia, Daniele Perito, and
  Abdelberi Chaabane.
\newblock {OMEN: Faster Password Guessing Using an Ordered Markov Enumerator}.
\newblock In {\em {International Symposium on Engineering Secure Software and
  Systems}}, milan, Italy, March 2015.

\bibitem{persuasion}
Alain Forget, Sonia Chiasson, P.~C. van Oorschot, and Robert Biddle.
\newblock {Improving Text Passwords through Persuasion}.
\newblock In {\em Proceedings of the 4th Symposium on Usable Privacy and
  Security}, SOUPS ’08, page 1–12, New York, NY, USA, 2008. Association for
  Computing Machinery.

\bibitem{ontheaccuracy}
Maximilian Golla and Markus D\"{u}rmuth.
\newblock {On the Accuracy of Password Strength Meters}.
\newblock In {\em Proceedings of the 2018 ACM SIGSAC Conference on Computer and
  Communications Security}, CCS ’18, page 1567–1582, New York, NY, USA,
  2018. Association for Computing Machinery.

\bibitem{hamme}
Daphne Koller and Nir Friedman.
\newblock {\em Probabilistic Graphical Models: Principles and Techniques -
  Adaptive Computation and Machine Learning}.
\newblock The MIT Press, 2009.

\bibitem{telepathwords}
Saranga Komanduri, Richard Shay, Lorrie~Faith Cranor, Cormac Herley, and Stuart
  Schechter.
\newblock {Telepathwords: Preventing Weak Passwords by Reading
  Users{\textquoteright} Minds}.
\newblock In {\em 23rd {USENIX} Security Symposium ({USENIX} Security 14)},
  pages 591--606, San Diego, CA, August 2014. {USENIX} Association.

\bibitem{pmlr-v31-ma13a}
Jianzhu Ma, Jian Peng, Sheng Wang, and Jinbo Xu.
\newblock {Estimating the Partition Function of Graphical Models Using Langevin
  Importance Sampling }.
\newblock In {\em Proceedings of the Sixteenth International Conference on
  Artificial Intelligence and Statistics}, volume~31 of {\em Proceedings of
  Machine Learning Research}, pages 433--441, Scottsdale, Arizona, USA, 29
  Apr--01 May 2013. PMLR.

\bibitem{gue_and_en}
J.~L. {Massey}.
\newblock {Guessing and entropy}.
\newblock In {\em Proceedings of 1994 IEEE International Symposium on
  Information Theory}, pages 204--, June 1994.

\bibitem{FLA}
William Melicher, Blase Ur, Sean~M. Segreti, Saranga Komanduri, Lujo Bauer,
  Nicolas Christin, and Lorrie~Faith Cranor.
\newblock {Fast, Lean, and Accurate: Modeling Password Guessability Using
  Neural Networks}.
\newblock In {\em 25th {USENIX} Security Symposium ({USENIX} Security 16)},
  pages 175--191, Austin, TX, August 2016. {USENIX} Association.

\bibitem{mm_first}
Arvind Narayanan and Vitaly Shmatikov.
\newblock {Fast Dictionary Attacks on Passwords Using Time-Space Tradeoff}.
\newblock In {\em Proceedings of the 12th ACM Conference on Computer and
  Communications Security}, CCS ’05, page 364–372, New York, NY, USA, 2005.
  Association for Computing Machinery.

\bibitem{IPGVRL}
Dario {Pasquini}, Ankit {Gangwal}, Giuseppe {Ateniese}, Massimo {Bernaschi},
  and Mauro {Conti}.
\newblock {{Improving Password Guessing via Representation Learning}}.
\newblock In {\em 2021 42th IEEE Symposium on Security and Privacy}, May 2021.

\bibitem{CAE}
Deepak Pathak, Philipp Kr\"ahenb\"uhl, Jeff Donahue, Trevor Darrell, and Alexei
  Efros.
\newblock {Context Encoders: Feature Learning by Inpainting}.
\newblock In {\em Computer Vision and Pattern Recognition ({CVPR})}, 2016.

\bibitem{FLA2}
Blase Ur, Felicia Alfieri, Maung Aung, Lujo Bauer, Nicolas Christin, Jessica
  Colnago, Lorrie~Faith Cranor, Henry Dixon, Pardis~Emami Naeini, Hana Habib,
  Noah Johnson, and William Melicher.
\newblock {Design and Evaluation of a Data-Driven Password Meter}.
\newblock In {\em CHI '17}, 2017.

\bibitem{dousers}
Blase Ur, Jonathan Bees, Sean~M. Segreti, Lujo Bauer, Nicolas Christin, and
  Lorrie~Faith Cranor.
\newblock Do users’ perceptions of password security match reality?
\newblock In {\em Proceedings of the 2016 CHI Conference on Human Factors in
  Computing Systems}, CHI ’16, page 3748–3760, New York, NY, USA, 2016.
  Association for Computing Machinery.

\bibitem{measure_up}
Blase Ur, Patrick~Gage Kelley, Saranga Komanduri, Joel Lee, Michael Maass,
  Michelle~L. Mazurek, Timothy Passaro, Richard Shay, Timothy Vidas, Lujo
  Bauer, Nicolas Christin, and Lorrie~Faith Cranor.
\newblock How does your password measure up? the effect of strength meters on
  password creation.
\newblock In {\em Presented as part of the 21st {USENIX} Security Symposium
  ({USENIX} Security 12)}, pages 65--80, Bellevue, WA, 2012. {USENIX}.

\bibitem{min-guess}
Blase Ur, Sean~M. Segreti, Lujo Bauer, Nicolas Christin, Lorrie~Faith Cranor,
  Saranga Komanduri, Darya Kurilova, Michelle~L. Mazurek, William Melicher, and
  Richard Shay.
\newblock {Measuring Real-World Accuracies and Biases in Modeling Password
  Guessability}.
\newblock In {\em 24th {USENIX} Security Symposium ({USENIX} Security 15)},
  pages 463--481, Washington, D.C., August 2015. {USENIX} Association.

\bibitem{fuzzyPSM}
D.~{Wang}, D.~{He}, H.~{Cheng}, and P.~{Wang}.
\newblock {fuzzyPSM: A New Password Strength Meter Using Fuzzy Probabilistic
  Context-Free Grammars}.
\newblock In {\em 2016 46th Annual IEEE/IFIP International Conference on
  Dependable Systems and Networks (DSN)}, pages 595--606, June 2016.

\bibitem{PCFG}
M.~{Weir}, S.~{Aggarwal}, B.~d.~{Medeiros}, and B.~{Glodek}.
\newblock {Password Cracking Using Probabilistic Context-Free Grammars}.
\newblock In {\em 2009 30th IEEE Symposium on Security and Privacy}, pages
  391--405, May 2009.

\bibitem{testing}
Matt Weir, Sudhir Aggarwal, Michael Collins, and Henry Stern.
\newblock {Testing Metrics for Password Creation Policies by Attacking Large
  Sets of Revealed Passwords}.
\newblock In {\em Proceedings of the 17th ACM Conference on Computer and
  Communications Security}, CCS ’10, page 162–175, New York, NY, USA, 2010.
  Association for Computing Machinery.

\bibitem{zxcvbn}
Daniel~Lowe Wheeler.
\newblock {zxcvbn: Low-Budget Password Strength Estimation}.
\newblock In {\em 25th {USENIX} Security Symposium ({USENIX} Security 16)},
  pages 157--173, Austin, TX, August 2016. {USENIX} Association.

\bibitem{inpainting}
Junyuan Xie, Linli Xu, and Enhong Chen.
\newblock Image denoising and inpainting with deep neural networks.
\newblock In {\em Advances in Neural Information Processing Systems 25}, pages
  341--349. Curran Associates, Inc., 2012.

\end{thebibliography}
